\documentclass[preprint,showpacs,preprintnumbers,amsmath,amssymb,aps]{revtex4-1}

\usepackage{times}
\usepackage{graphicx,dcolumn,bm}
\usepackage{amssymb,amstext,amsmath}
\usepackage{graphicx}
\usepackage{dcolumn}
\usepackage{bm}
\usepackage{amssymb}
\usepackage{multirow}
\usepackage{bigstrut}
\usepackage{makecell,rotating}
\usepackage{mathrsfs}
\usepackage{booktabs}
\usepackage{threeparttable}
\usepackage{multirow}
\usepackage{subfigure}
\usepackage{epsfig}
\usepackage{threeparttable}
\usepackage{chngpage}
\usepackage{float}
\usepackage{amstext}
\usepackage{amsmath}
\usepackage{setspace}

\usepackage{color}
	 \definecolor{darkred}{rgb}{0.75,0,0}
	 \definecolor{darkgreen}{rgb}{0,0.5,0}
	 \definecolor{darkblue}{rgb}{0,0,0}
  	 \definecolor{darkorange}{rgb}{1,0.9,0.1}
\newcommand{\aming}[1]{\textcolor{darkblue}{#1}}

\usepackage[top=23.4mm, bottom=23.4mm, left=23.4mm,
right=23.4mm]{geometry}

\newcommand{\cee}{\mathbf{c}}
\newcommand{\x}{\mathbf{x}}
\newcommand{\A}{\mathbf{A}}
\newcommand{\Q}{\mathbf{Q}}
\newcommand{\B}{\mathbf{B}}
\newcommand{\R}{\mathbf{R}}
\newcommand{\W}{\mathbf{W}}
\newcommand{\N}{\mathbf{N}}
\newcommand{\h}{\mathbf{H}}
\newcommand{\K}{\mathbf{K}}
\newcommand{\U}{\mathbf{U}}
\newcommand{\Weff}{\mathbf{W}_{\textrm{eff}}}
\newcommand{\zero}{\mathbf{0}}
\newcommand{\z}{\mathbf{z}}
\newcommand{\E}{E}
\newcommand{\eig}{\lambda_\textrm{min}}
\renewcommand{\v}{\mathbf{v}}
\renewcommand{\u}{\mathbf{u}}
\renewcommand{\d}{\mathbf{d}}
\newcommand{\s}{\mathbf{S}}
\newcommand{\e}{\mathrm{e}}
\newcommand{\dt}{\Delta t}

\newcommand{\ie}{\textit{i.e.}}

\renewcommand{\thefootnote}{\Alph{footnote}}
\newcommand{\Eavg}{\langle \E \rangle}
\newcommand{\Eub}{\overline{\E}}
\newcommand{\Elb}{\underline{\E}}

\begin{document} 

\title{The fundamental advantages of temporal networks} 

\author{Aming Li,$^{1,2}$ Sean P. Cornelius,$^{1,3}$ Yang-Yu Liu,$^{3,4}$ Long Wang,$^{2,\ast}$ Albert-L\'{a}szl\'{o} Barab\'{a}si$^{1,4,5,6,}$}
\email{Corresponding authors.}

\affiliation{$^{1}$Center for Complex Network Research and Department of Physics,
Northeastern University, Boston, MA 02115, USA\\
$^{2}$Center for Systems and Control, College of Engineering,
Peking University, Beijing 100871, China\\
$^{3}$Channing Division of Network Medicine, Brigham and Women's Hospital,
Harvard Medical School, Boston, MA 02115, USA\\
$^{4}$Center for Cancer Systems Biology, Dana-Farber Cancer Institute,
Boston, MA 02115, USA\\
$^{5}$Department of Medicine, Brigham and Women's Hospital, Harvard Medical School,
Boston, MA 02115, USA\\
$^{6}$Center for Network Science, Central European University, Budapest 1052, Hungary
}

\date{\today}

\begin{abstract}
Despite the traditional focus of network science on static networks, most networked systems of scientific interest are characterized by temporal links.
By disrupting the paths, link temporality has been shown to frustrate many dynamical processes on networks, from information spreading to accessibility.
Considering the ubiquity of temporal networks in nature, we must ask: Are there any advantages of the networks' temporality?
Here we develop an analytical framework to explore the control properties of temporal networks, 
arriving at the counterintuitive conclusion that 
temporal networks, compared to their static (\ie~aggregated) counterparts, reach controllability faster,
demand orders of magnitude less control energy, and the control trajectories, through which the system reaches its final states, are significantly more compact than those characterizing their static counterparts. 
The combination of analytical, numerical and empirical results demonstrates that temporality ensures a degree of flexibility that would be unattainable in static networks, significantly enhancing our ability to control them. 
\end{abstract}

\maketitle

\tableofcontents 

\newpage
\section{Introduction}
Traditionally network science has focused on static networks, whose links offer permanent connections between their nodes.
Yet, it is increasingly recognized that most natural and social systems are best described as \emph{temporal} networks, acknowledging that their links exist only intermittently. 
For example, within cellular metabolic networks \cite{Almaas2004}, the links correspond to brief chemical reactions; 
in social networks \cite{Caldarelli2007,Havlin2004book,Goncalves2015book,Toroczkai2007,Newcomb1961,Snijders2001} friendship links are inferred from face-to-face or digital communications of short duration. 
\aming{A substantial body of research has found that} 
such temporality \cite{Slowingd13PRL,Accessibility13PRL,Causality14NatCom,Lawyer15SR,Holme2015review} has profound effects on most dynamical processes taking place on networks \cite{Causality14NatCom,Slowingdon12PRE,Slowingd13PRL,Accessibility13PRL}, slowing down synchronization and the diffusion of innovative information \cite{Slowingd13PRL}, impeding exploration and navigation \cite{Slowingdon12PRE}, and raising barriers to accessibility \cite{Accessibility13PRL}.
\aming{Similar limitations are expected for control---the ability to drive a system with input signals to any desired final state in finite time. 
Control is essential for the operation of most real systems  \cite{Kalman63,ControlinRealSysTac04,ConRealSysCell11,ConRealSysPNAS12,Nepusz2012,Chen2014}, yet controllability normally
requires the existence of continuous paths capable of carrying the input signals to the rest of the network  \cite{Liu2011,Rajapakse2011pnas,Ruths2014science}.}
While in static networks such signal-carrying paths are permanently available, in temporal systems complete instantaneous paths between the inputs and the rest of the nodes are not guaranteed, potentially degrading our ability to control a system.

\section{Dynamics on temporal networks}
A temporal network is an ordered sequence of $m = 1,\cdots,M$ separate networks on the same set of $N$ nodes (Fig.~\ref{fig_main_1}A and~\ref{fig_main_1}B), with each such \emph{snapshot} $m$
characterized by a (weighted) adjacency matrix $\A_m$ for a duration $\tau_m$.
As we aim to uncover the role of the changing network topology, rather than the effect of specific dynamics, we consider that in each snapshot the system is governed by the canonical linear time-invariant dynamics,
\begin{equation}
\label{temporaldynamics}
\dot{\x}(t)=\A_m \x(t) + \B_m \u_m(t),
\end{equation}
valid over the time interval $t \in [t_{m-1}, t_{m-1}+\tau_m)$, where the $t_{m-1}=\sum_{j=1}^{m-1}\tau_j$ is the \textit{switching time} between snapshots $m-1$ and $m$; 
$x_i(t)$ represents the state of each node $i$ at time $t$, like the concentration of metabolite $i$ within a cell;
the input matrix $\B_m$ identifies the set of driver nodes through which we attempt to control the system using $p$ independent control inputs $\u_m(t)\in\mathbb{R}^p$, like manipulating the input concentration of $p$ metabolites.  
To avoid conferring an unfair advantage to temporal networks, we use the same set of driver nodes across all snapshots, \ie~$\B_m = \B$ (Fig.~\ref{fig_main_1}C), and 
we assume that we have no control over the order of the snapshots, nor over the timings of the topology changes, 
hence our influence on the system is confined to the control inputs
 \cite{Liu2011,Rajapakse2011pnas,Ruths2014science,Posfai2014NJP,Yangreview15}.
Although the dynamics Eq.~(\ref{temporaldynamics}) within each snapshot is linear, the switching process makes the overall dynamics nonlinear \cite{Guo2003}. 
Indeed, temporal networks with dynamics described by Eq.~(\ref{temporaldynamics}) are a special case of switched systems \cite{Xie2002tac}, which can exhibit exotic collective behavior absent in linear systems like multiple limit cycles, chaos, and the Zeno-like phenomenon \cite{ChaseTAC93,Savkin01Au}.

\section{Time to control}
\begin{figure}[H]
\centering
\includegraphics[width=\textwidth]{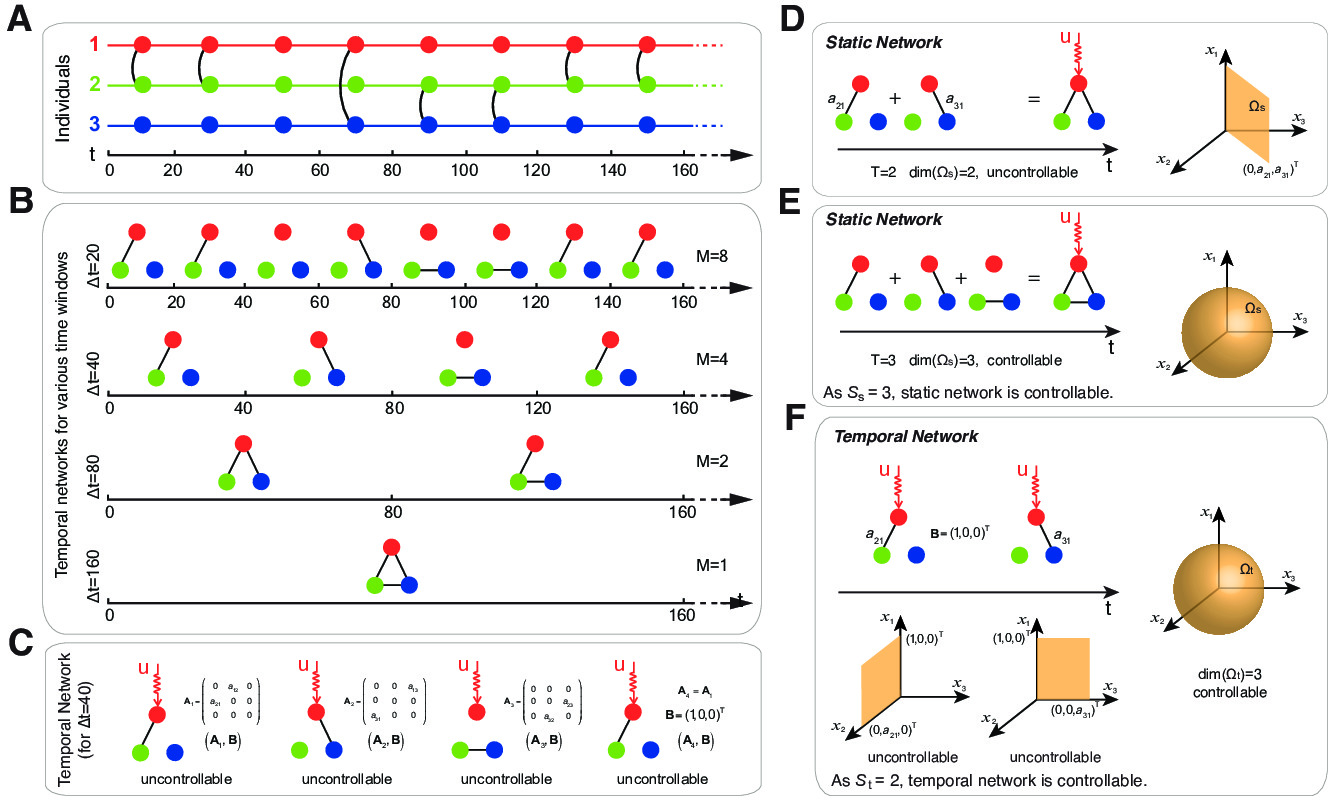}
\caption{
\textbf{Temporal vs. static networks.}
(\textbf{A}) The sequence of contacts among three nodes, capturing, for example, the connection patterns between three individuals.
We draw a line between two nodes if they interact with each other during a $20$ second interval.
(\textbf{B}) The temporal networks constructed from the contact sequence shown in (A) depend on the length of the time window $\dt$ over which we aggregate the interactions.
For short $\dt$ we obtain a large number of disconnected networks (snapshots); for sufficiently large $\dt$ these collapse into a single static network.
(\textbf{C}) Four snapshots of a temporal network, constructed from 
(A) for $\dt = 40$.
According to Kalman's rank condition \cite{Kalman63}, none of the snapshots is individually controllable from the top node.
Yet, Eq.~(\ref{theorm01}) predicts
that the temporal network becomes controllable after the second snapshot.
(\textbf{D}) A static network obtained by aggregating the first two snapshots in (C), is uncontrollable from the top node, as the controllable space $\Omega_\text{s}$ is two dimensional.
This is illustrated by the yellow region on the right panel showing the set of all points $\x_f$ that can be reached in finite time from a given initial state $\zero$ with an appropriate $\u(t)$. 
(\textbf{E}) We must aggregate at least three ($S_\text{s}=3$) for the corresponding static network to become controllable.
(\textbf{F}) The first two snapshots of a temporal network (top) shown in (C) and the controllable space for each snapshot (bottom).
In this case the temporal network becomes controllable after $S_\text{t}=2$ snapshots.
The controllable space of the temporal network is shown by the sphere on the right panel.
}
\label{fig_main_1}
\end{figure}

To understand why temporal networks are easier to control than static networks, consider the static three-node network of Fig.~\ref{fig_main_1}D, which
is uncontrollable by any single driver node.
Indeed, we can only steer the network within a two-dimensional subspace of the three-dimensional state space (Fig.~\ref{fig_main_1}D). 
Yet, the temporal version of the same network---in which the two links are non-simultaneously active---\emph{is} controllable by the top node (Fig.~\ref{fig_main_1}F),
as each snapshot of the temporal network \aming{contributes} an \emph{independent} two-dimensional controllable space, making the full three-dimensional state space accessible. 
 
To generalize these concepts to an arbitrary temporal network, consider a system initially at $\x_0 = \zero$. 
By considering all possible trajectories from $t_0=0$ to $t_f=t_M$, we can write the controllable space 
(see SI) 
 as
\begin{equation}
\label{controllableset_main}
\Omega = \langle \A_M | \B \rangle + \sum_{m=1}^{M-1} \prod_{j=M}^{m+1} \e^{\A_j \tau_j}  \langle \A_m | \B \rangle.
\end{equation}
Here $\langle \A_m | \B \rangle = \sum_{i=0}^{N-1}\A^{i}_m\R(\B)$ denotes the controllable space of snapshot $m$, where $\R(\B)=\{\B \v | \v \in \mathbb{R}^p\}$ is
the column space of $\B$.
Hence, a temporal network of $N$ nodes is controllable if and only if
\begin{equation}
\label{theorm01}
\Omega = \mathbb{R}^N,
\end{equation}
meaning that we can steer the system to an arbitrary state of the state space $\mathbb{R}^N$.
For a static network, whose snapshots are identical (\aming{\ie}~$\A_m = \A$), Eq.~(\ref{theorm01}) reduces to the classic Kalman rank condition \cite{Kalman63} for controllability.

According to Eq.~(\ref{controllableset_main}), the controllable space grows as the network structure changes,
allowing us to determine the number of snapshots $S_\text{t}$ that must elapse before a temporal network becomes fully controllable.
For comparison, we also calculate the number of snapshots $S_\text{s}$ we must \emph{aggregate} 
in order to obtain a controllable static network 
(see SI) under the same sequence of snapshots and set of driver nodes. 
\aming{Consider for example} Fig.~\ref{fig_main_1}C, which shows a temporal network \aming{with} four snapshots, none of which is individually controllable using the top node as the sole driver node.
According to Eq.~(\ref{theorm01}) the temporal sequence becomes controllable at the second snapshot, \ie~$S_\text{t}=2$ (Fig.~\ref{fig_main_1}F).
In contrast, we must aggregate $S_\text{s}=3$ snapshots to obtain a controllable static network (Fig.~\ref{fig_main_1}D, E). 
The difference between $S_\text{s}$ and $S_\text{t}$ captures the relative control benefits of a dense (but fixed) network topology versus a relatively sparse (but time-varying) topology, respectively.
While there is no theoretical guarantee that $S_\text{t}$ is always less than $S_\text{s}$, 
as we show next, we find that real temporal networks reach controllability much faster
than their static counterparts. 
Here ``faster'' refers to the number of snapshots we need to reach full controllability.
Hence the time to control (embodied in $S_\text{t}$ and $S_\text{s}$) is distinct from $t_f$, representing the time a system
needs to reach its final state \cite{Kalman63,Xie2002tac,Liu2011,Sun2013prl,Yan2015a}.

To demonstrate the practical relevance of our finding, we explore 
an empirical communication dataset collected by the SocioPatterns collaboration, capturing face-to-face conversations between the attendees of a conference \cite{Isella2011}; 
an ecological network, capturing antenna-body interactions between ants \cite{BlonderAnt2011};
a biological network capturing the temporal dynamics of protein-protein interactions \cite{PPIdata};
and a technological network recording data packet exchanges in an emulated mobile ad-hoc network.
The aggregation window $\dt$ (over which we condense interactions into snapshots) offers an inverse measure of temporality: for small $\dt$, we obtain a large number of 
sparse, disjoint snapshots, while for large $\dt$ we approach a single snapshot corresponding to the full static network (Fig.~\ref{fig_main_1}A and~\ref{fig_main_1}B). 
Figure~\ref{fig_main_2} shows $S_\text{t}$ and $S_\text{s}$ for different time windows $\dt$, indicating that the number of snapshots required to control a temporal network $S_\text{t}$ is always less than that for its static counterpart, $S_\text{s}$.
The higher the temporality (smaller $\dt$), the more pronounced the advantage of temporal networks.
For example, for the face-to-face interactions at $\dt = 10^3$ the temporal networks reach controllability after
only $71$ time steps (19.72 hours), while we must aggregate $185$ snapshots over two days to 
obtain a static network that is controllable.
To see to what extent these gains in controllability depend on the topology of the underlying snapshots, or on the inter-event characteristics of the temporal patterns, we calculate $S_\text{t}$ and $S_\text{s}$ for randomized versions of the empirical data, using null models that permute the overall network structure, the relative time orderings of the edges, 
or both \cite{Holme2012}. 
In all cases, we find that $S_\text{t} < S_\text{s}$ (Fig.~\ref{fig_main_2}), indicating that the observed advantage is independent of how the network changes in time.
In other words, temporality alone is sufficient to improve controllability.

\section{Control energy} 
Our ability to control a system is determined not only by the time it takes to reach controllability, but also by the amount of effort (energy) required to reach a particular final state. 
We develop a formalism (see SI) to calculate the minimum input energy $\frac{1}{2}\int_{t_0}^{t_f}\u^\mathrm{T}(t)\u(t)\mathrm{d}t$ required to drive a temporal network from an initial state $\x_0$ to final state $\x_f$, obtaining
\begin{equation}
\label{retemp}
\E(\x_0, \x_f)  =  \frac{1}{2} \d^\mathrm{T} \W_{\textrm{eff}} ^{-1} \d,
\end{equation}
where $\d$ is the difference vector between the desired final state $\x_f$ and the natural final state that the system reaches  without control inputs, and the  $N \times N$ matrix $\Weff$ encodes the energy structure of the temporal network. 
For identical snapshots $\Weff$ reduces to the controllability gramian $\W$ and Eq.~(\ref{retemp}) provides the control energy of a static network 
(see SI).

Figure~\ref{fig_main_3} compares the control energy Eq.~(\ref{retemp}) of a typical synthetic temporal network as well as the technological network with its static counterpart for various aggregation times $\dt$ (see SI).
In both cases we find that the average control energy of a temporal network is many orders of magnitude smaller than that of the corresponding static network.
The effect is particularly pronounced for small $\dt$: for $\dt = 10^{-6}$ the energy difference between the static and temporal network exceeds $130$ orders of magnitude (Fig.~\ref{fig_main_3}A, B).
In other words, high temporality, corresponding to rapid changes in the network topology, offers remarkable energy savings. 
For example, the energy required to control the technological network of Fig.~\ref{fig_main_2}D 
drops by 274 orders of magnitude when the true temporal nature of the network is taken into account
(Fig.~\ref{fig_main_3}D).
In both temporal and static networks we can reduce the control energy by using more driver nodes \cite{Yan2015a}
(Fig.~\ref{fig_main_3}B).
Yet the gap between the temporal and static network persists until \emph{all} nodes are directly controlled (Fig.~\ref{fig_main_3}C, E). 
Finally, a direct comparison of the control energy distributions for static vs. temporal networks reveals only a small overlap in certain regimes (Fig.~\ref{fig_main_3}, insets). 
This indicates that the \emph{worst}-case control direction in a temporal network is often better than the \emph{best} case control direction in its static counterpart.

The extreme energy savings characterizing temporal networks arise from the fact that there are orders of magnitude differences in the energy required to move in different directions in the state space \cite{Yan2015a}. 
In a static network, if we must travel in an energetically costly direction, we have no choice but to spend the required energy to ``push'' against the system's dynamics.
By contrast, in a temporal network we can exploit the changing topology to avoid these expensive directions. 
Much like navigating a sailboat, it is easier to travel in a particular direction if we exploit the shifts in the wind direction (the vector field of the network dynamics):
we raise the sails when the wind helps us and pull them back when it works against us.
In other words, we only push towards the desired final state when the topology renders the energy cost acceptable, and pause when the topology 
makes the cost prohibitive.

\section{Locality of the control trajectories}
Real systems often obey constraints that forbid the states of the nodes from taking arbitrary values. 
For instance, the generator frequencies in the power grid can only vary within a narrow range around their normal 
operating point, without inducing failures, and metabolite concentrations \cite{Almaas2004} within a cell must always be non-negative.
These limitations mean that the control trajectories cannot wander arbitrarily far into the state space, but must exhibit a high degree of \emph{locality} \cite{Sun2013prl}.

To test the degree of locality in temporal networks, we calculate the length of each control trajectory using $L = \int_{t_0}^{t_f} \| \dot{\x}(t) \| \mathrm{d}t$,
where the energy-optimal trajectory for temporal networks as they move from $\x_0$ to $\x_f$ follows
\begin{equation*}
\label{opti_traj}
\x(t) = \e^{\A_m (t - t_{m-1})} \x(t_{m-1}) + \textrm{\textbf{W}}_m[t_{m-1}, t]\cee^*_m
\end{equation*}
for $t \in [t_{m-1}, t_{m-1}+\tau_m)$, where $\textrm{\textbf{W}}_m[t_{m-1}, t]$ is the gramian matrix of snapshot $m$, and $\cee^*_m$ is a constant vector of dimension $N$ 
(see SI).
In general, $L$ depends on the initial state $\x_0$ and the state space distance $\delta = \| \x_f - \x_0 \|$ between the origin and destination points (Fig.~\ref{fig_main_4}B).
For $\x_0=\zero$, in both static and temporal networks, $L$ increases linearly with $\delta$ (see Fig.~\ref{fig_main_4}B and Figs.~S13, S14, S16).
Yet, for any $\delta$ the optimal control trajectories in temporal networks are about five orders of magnitude shorter than in their static counterparts (Fig.~\ref{fig_main_4}B).
The difference is particularly remarkable in real systems: for the technological network the temporal trajectory is $29$ orders of magnitude shorter than for the corresponding static network (Fig.~\ref{fig_main_4}C).
Indeed, it is known that in static networks $L$ can be large even when 
$\delta$
approaches zero, implying static control generally demands highly non-local control trajectories \cite{Sun2013prl}.
In contrast, we find that the dynamical flexibility offered by temporality allows $\x(t)$ not to have to wander far into phase space as it moves from $\x_0$ to $\x_f$.
The sailing analogy helps once again to understand this difference: 
Travel against the prevailing wind is possible, but we must use complicated zig--zagging 
maneuvers
to reach our destination.  
If, however, the wind occasionally changes directions, a strategic use of the sails allows us to 
travel more directly to our destination.

\section{Discussion} 
Previous studies have shown that static networks, even if they theoretically satisfy the controllability conditions, can be difficult to control in practice. 
For example, networks with heterogeneous degree distributions, ubiquitous in natural systems, require a high fraction of the nodes to be directly controlled \cite{Liu2011}. 
Making matters worse, the prohibitive energetic cost of some trajectories leads to an unavoidable tradeoff between the number of driver nodes and the time to control, and the nonlocality of the available control trajectories \cite{Sun2013prl,Yan2015a} 
further strains the controllability of static systems.
How do we reconcile this practical difficulty of control with the observation that 
in order to perform their proper functions, real natural and technological systems must have the ability to drive themselves to (and keep themselves in) desired states?
Temporal networks appear to be a promising solution to this conundrum. 
When we aim to control a network that periodically changes its structure, we observe \emph{simultaneous} reductions in all relevant control metrics, from time to control to control energy and trajectory lengths, the effect size surpassing many orders of magnitude.
Therefore, our results suggest a hitherto unappreciated role for temporality, which has traditionally been viewed as an impediment for network dynamics. 
By exploiting the best parts of the dynamics of each subsnapshot, we can therefore enjoy the ``best of all worlds.''

Finally, the temporal networks we explored
are inherently nonlinear systems.
Consistent with previous finding that nonlinearity can help control dynamical systems, from driving systems to desired attractors \cite{Cornelius2013NatCommun} to chaos control \cite{Cavalcante13PRL}, our results indicate that nonlinearity induced by temporality can be an asset, rather than an obstacle\aming{, to control}.

\noindent \textbf{Acknowledgments:}
We thank G. Xie, Y. Guan, G. Yan, I. Kov\'{a}cs, I. Scholtes, M. Angulo, E. Guney, B. Barzel, J. Gao, R. Sinatra, S. Mou, Z. Ji, J. Huang, K. Albrecht, M. Santolini, M. Szell, I. Donnelly, R. Rajaei, S. Aleva, B. Common, and J. De Nicolo for valuable discussions and help.
We appreciate J. Wang and X. Peng for supplying the data of dynamic protein-protein interaction networks.
We wish to thank M. Gavin and the DARPA Project W911NF-12-C-002 for providing the mobile network datasets.
This work is supported by Army Research Laboratories (ARL) Network Science (NS) Collaborative Technology Alliance (CTA) grant ARL NS-CTA W911NF-09-2-0053; 
the John Templeton Foundation: Mathematical and Physical Sciences grant number PFI-777; European Commission grant numbers FP7 317532 (MULTIPLEX) and 641191 (CIMPLEX).
A.L. and L.W. are supported by NSFC (Grants No. 61375120 and No. 61533001).
A.L. also acknowledges the support from China Scholarship Council (No. 201406010195).

\noindent \textbf{Author Contributions:}
A.-L.B. and S.P.C. conceived the project.  
All authors designed and performed the research, and analyzed the results.
A.L. and L.W. performed all the analytical calculations. 
A.L. performed all the numerical calculations (except the technology network that was analyzed by Y.-Y.L. independently). 
A.-L.B., A.L., and S.P.C. led the writing of the manuscript,
Y.-Y.L. and L.W. edited the manuscript.

\newpage

\begin{figure}[H]
\centering
\includegraphics[width=\textwidth]{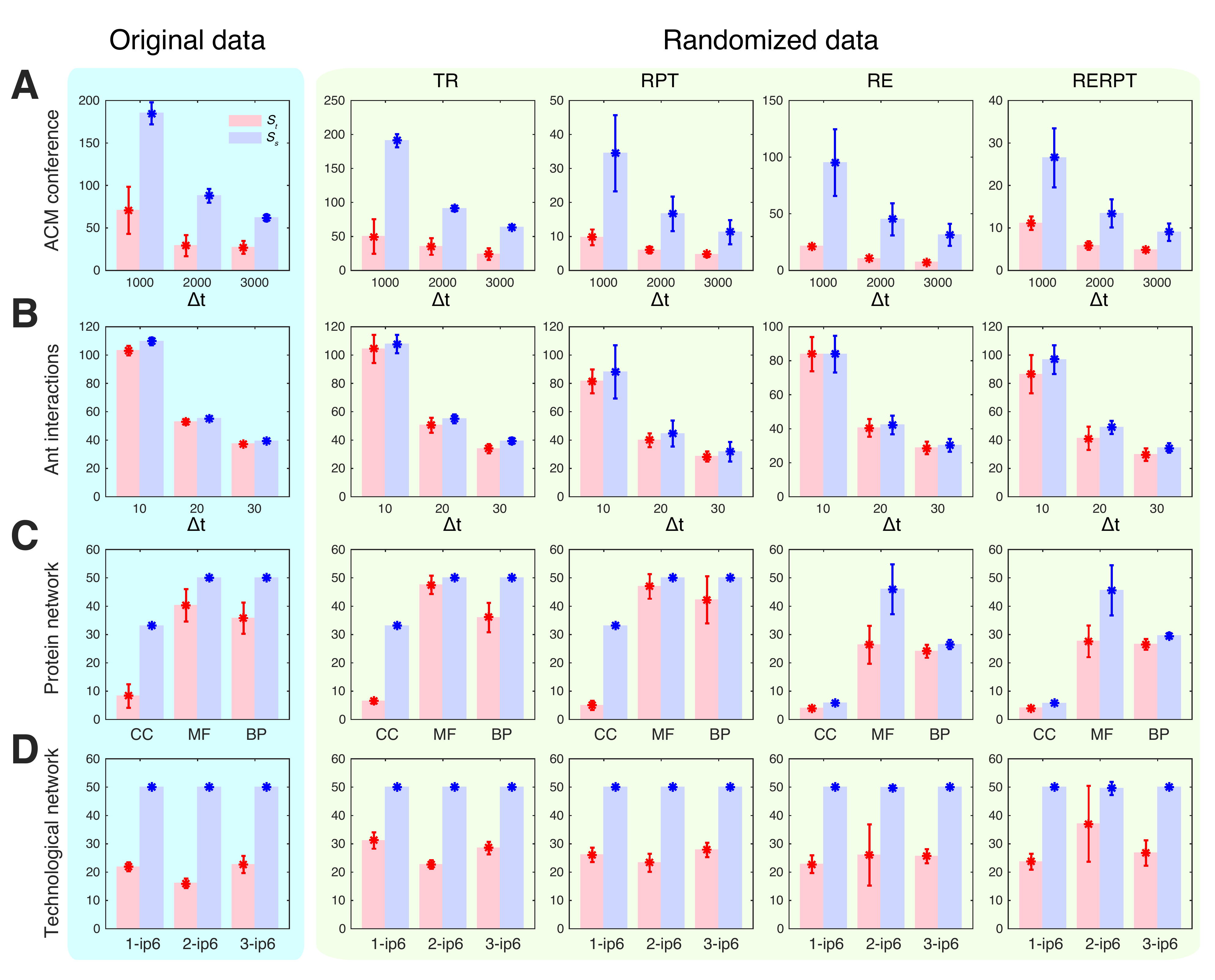}
\caption{
\textbf{Faster paths to controllability in temporal networks.}
Time to control in four kinds of temporal networks:
(\textbf{A}) Interactions between the attendees of an ACM hypertext conference of 113 participants, recorded over 2.5 days in 2009 \cite{Isella2011};
(\textbf{B}) interactions between 89 ants over 1,438 seconds \cite{BlonderAnt2011};
(\textbf{C}) dynamic protein-protein interactions (PPI) in \emph{Saccharomyces cerevisiae}, annotated by the three domains of Gene Ontology: cellular component (CC), molecular function (MF), and biological process (BP) 
(see SI). 
The resulting three temporal networks have 33, 50, 50 snapshots with 84, 74, 85 nodes, respectively \cite{PPIdata};
\textbf{(D)} data packet exchanges in an emulated mobile ad-hoc network with 50 snapshots and 34 nodes (datasets 1-ip6, 2-ip6, and 3-ip6, see 
SI
for details).
For each time window $\dt$,
$S_\text{t}$ is the minimum number of snapshots required to achieve full control of a temporal network,
and $S_\text{s}$ is the minimum number of snapshots we must aggregate to obtain a controllable static network.
We find that $S_\text{t}<S_\text{s}$ for any $\dt$ both for the original sequence of snapshots, and when the sequence of interactions is randomized using several null models \cite{Holme2012}: TR (Time Reversal), RPT (Randomly Permuted Times), RE (Randomized Edges) and RERPT (Randomized Edges and Randomly Permuted Times) (see 
SI
for the randomization procedure).
The weight of each link is set randomly between $0$ and $1$. 
The number of driver nodes is fixed to $20\%$ of the network size, and each point is averaged over $10^3$ realizations of link weights on the same network structure (temporal or static).
}
\label{fig_main_2}
\end{figure}

\newpage
\begin{figure}[H]
\centering
\includegraphics[width=\textwidth]{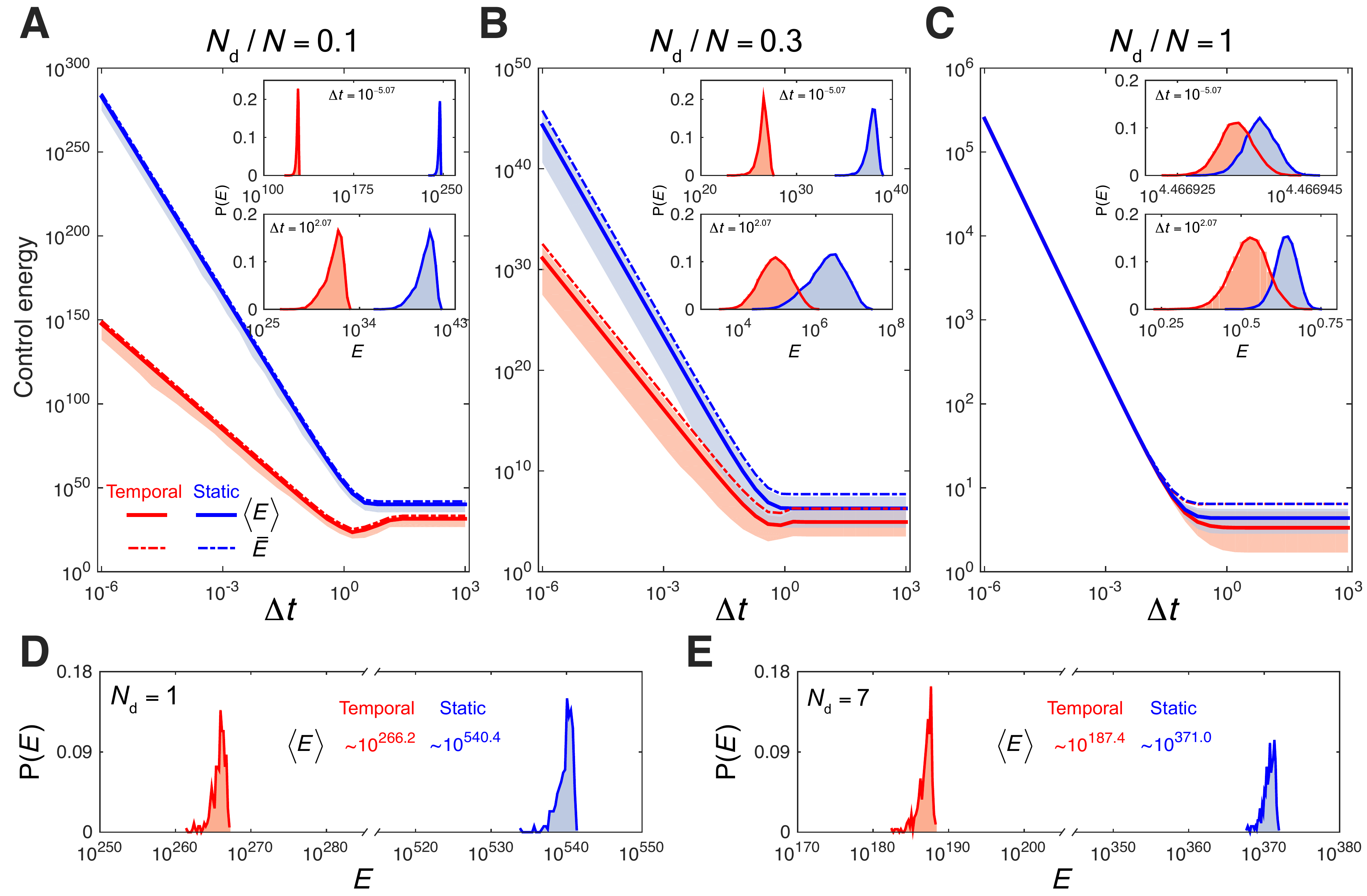}
\caption{
\textbf{Temporal networks require less control energy.}
Control energy for temporal and static networks.
The fractions of nodes controlled in the three panels are: \textbf{(A)} $0.1$, \textbf{(B)} $0.3$, and \textbf{(C)} $1$.
We consider a temporal network with two snapshots, each lasting $\dt$ time.
$\Eavg$ with solid line is the average minimum energy over $10^4$ randomly selected final states $\x_f$ from a unit sphere centered on $\x_0 = \zero$. 
Each shaded area is enclosed by the minimum and maximum $E$ obtained numerically. 
$\Eub$ with dashed line is the theoretical upper bound of $E$ 
(see SI).
With increasing $\dt$, $\Eub$ (and $\Eavg$) decreases as $\Eub \sim \dt ^{-\gamma} $, before reaching a plateau. 
Increasing the number of driver nodes reduces the necessary control energy both for temporal and static networks, yet temporal networks continue to require less energy.
Insets: the distribution of minimal energy for two specific $\dt$.
The small overlap between temporal and static networks
indicates that temporal networks are energetically preferable and largely independent of the control direction.
Here $a_1 = -3$, $a_2 = -1$, $\bar{k} = 6$, $N = 20$.
For the technological network with $\dt=10^{-6}$, \textbf{(D)} and \textbf{(E)} show the distribution of $E$ for $N_d = 1$ and $7$, respectively (see Fig.~S12 for more details).
 }
 \label{fig_main_3}
\end{figure}

\newpage
\begin{figure}[H]
\includegraphics[width=1\textwidth]{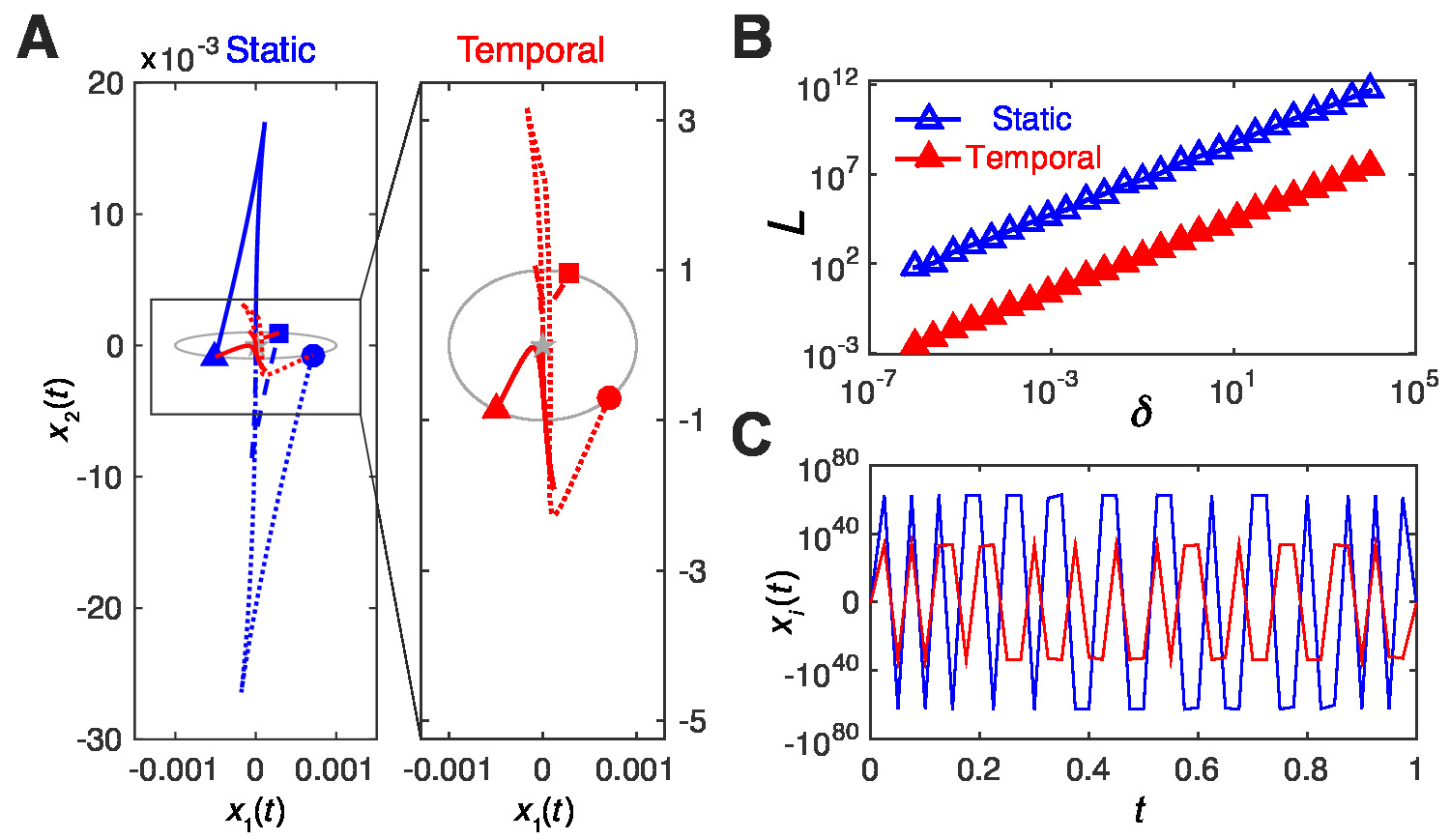}
\caption{
\textbf{Temporal networks exhibit more local trajectories.}
(\textbf{A}) Three trajectories for static (blue lines) and temporal (red lines) networks starting from $\x_0 = \zero$, controlled to reach $\| \x_f \| = 10^{-3}$ after a unit time.
The blow up of the control region, shown on the right, allows us to better observe the more localized temporal control trajectories.
(\textbf{B}) The length of the control trajectory $L$ in function of control distance $\delta = \| \x_f - \x_0 \|$.
$L$ is always much smaller for the temporal networks than in their static counterparts, regardless of the control distance (see Fig.~S14).
Each point represents an average over $10^4$ final states;
we choose $N=10, M=5$.
See also Figs. S13, S14, S15, and S17 for other parameters.
(\textbf{C}) For the technological network, we track the evolution of $x_1(t)$ (corresponding to the maximum length of all state components) from $x_1(0)=0$ to $x_1(1) = 1/\sqrt{34}$, finding that $L$ is in the order of $10^{35}$ for the temporal network, in contrast with $10^{64}$ for the corresponding static network (see Fig.~S18 for more details).
}
\label{fig_main_4}
\end{figure}

\newpage
\setcounter{figure}{0}
\renewcommand{\thefigure}{S\arabic{figure}}


\appendix
\section{Controllable space}
\label{ConctrollanleStateSet}
\subsection{Temporal networks}
\label{SI_temConSpa}

For a dynamical system of the form
\begin{eqnarray*}
\dot{\x }(t) & = & \A \x (t)+\B\u(t),
\end{eqnarray*}
we can write the system state at time $t$ as
\begin{eqnarray*}
\x (t) & = & \e^{\A (t-t_0)}\x _0 + \int^t_{t_0} \e^{\A (t-\tau)}\B \u(\tau) \textrm{d} \tau,
\end{eqnarray*}
with the initial state $\x (t_0) = \x _0$.
Hence, for temporal networks we can write the system's state at the switching times as
\begin{eqnarray*}
\x (t_1) & = & \e^{\A _1\tau_1}\x _0 +\int_{t_{0}}^{t_1}\e^{\A _{1}(t_1-s)}\B _{1}\u_1(s)\mathrm{d}s
\\ \nonumber
& \vdots &
\\ \nonumber
\x (t_m)& = & \e^{\A _m\tau_m}\x (t_{m-1}) +\int_{t_{m-1}}^{t_m}\e^{\A _{m}(t_m-s)}\B _{m}\u_m(s)\mathrm{d}s
\\ \nonumber
& \vdots &
\\ \nonumber
\x (t_M) & = & \e^{\A _M\tau_M}\x (t_{M-1}) +\int_{t_{M-1}}^{t_M}\e^{\A _{M}(t_M-s)}\B _{M}\u_M(s)\mathrm{d}s = \x _f.
\end{eqnarray*}
Thus after $M$ snapshots, the final state $\x _f$ at time $t_M$ is
\begin{eqnarray}
\label{solutionfortem}
\x _f &=& \prod_{m=M}^1 \e^{\A _{m}\tau_m}\x _0 + \sum_{m=1}^{M-1} \left(\prod_{j=M}^{m+1}\e^{\A _{j}\tau_j}\int_{t_{m-1}}^{t_m}\e^{\A _{m}(t_m-s)}\B _{m}\u_m(s)\mathrm{d}s\right) \\ \nonumber
   & & + \int_{t_{M-1}}^{t_M}\e^{\A _{M}(t_M-s)}\B _{M}\u_M(s)\mathrm{d}s.
\end{eqnarray}
 
Hence, we can write all states reachable from $\x _0=0$ as
\begin{eqnarray*}
\x _f &=& \sum_{m=1}^{M-1} \left(\prod_{j=M}^{m+1}\e^{\A _{j}\tau_j}\int_{t_{m-1}}^{t_m}\e^{\A _{m}(t_m-s)}\B _{m}\u_m(s)\mathrm{d}s\right)  
+ \int_{t_{M-1}}^{t_M}\e^{\A _{M}(t_M-s)}\B _{M}\u_M(s)\mathrm{d}s.
\end{eqnarray*}
Similarly, we can write all states that can reach $\x _f=0$ as:
\begin{eqnarray*}
\label{x0}
\x _0 &=& - \sum_{m=1}^M \prod_{j=1}^m \e^{-\A _{j}\tau_j} \int_{t_{m-1}}^{t_m}\e^{\A _{m}(t_m-s)}\B _{m}\u_m(s)\mathrm{d}s.
\end{eqnarray*}
Taken together, the set of controllable states for a temporal network defined by $\{(\A _{m}, \B_m , \tau_m)\}^M_{m=1}$ is
\begin{eqnarray*}
\Omega &=& 
\sum_{m=1}^{M-1} \prod_{j=M}^{m+1}\e^{\A _{j}\tau_j}
\left \{\x \bigg| \x = \int_{t_{m-1}}^{t_m}\e^{\A _{m}(t_m-s)}\B _{m}\u_m(s)\mathrm{d}s, \text{~for}~\forall \u_m \right\}
\\ \nonumber & &
+\left \{\x \bigg| \x = \int_{t_{M-1}}^{t_M}\e^{\A _{M}(t_M-s)}\B _{M}\u_M(s)\mathrm{d}s, \text{~for}~\forall \u_M \right\},
\end{eqnarray*}
and the corresponding set of reachable states is
\begin{eqnarray*}
\sum_{m=1}^M \prod_{j=1}^m \e^{-\A _{j}\tau_j}  \left \{\x \bigg| \x = \int_{t_{m-1}}^{t_m}\e^{\A _{m}(t_m-s)}\B _{m}\u_m(s)\mathrm{d}s, \text{~for}~\forall \u_m \right\}.
\end{eqnarray*}

For the integration part of the above expression, we have a lemma. \\
\textit{\textbf{Lemma}}. Given matrices $\A  \in \mathbb{R}^{N\times N}$ and $\B  \in  \mathbb{R}^{N\times p}$, for any $0 \leq t_0 < t_f < +\infty$, we have
\begin{equation}
\label{lemma1}
\left \{\x \bigg| \x = \int_{t_0}^{t_f}\e^{\A  (t_f-s)}\B  \u(s)\mathrm{d}s, \forall \u \right\}=\langle \A |\B  \rangle,
\end{equation}
where
\begin{equation*}
\langle \A  | \B  \rangle=\langle \A  | \mathrm{\textbf{R}}(\B ) \rangle=\mathrm{\textbf{R}}(\B ) + \A \mathrm{\textbf{R}}(\B ) +\cdots+\A ^{N-1}\mathrm{\textbf{R}}(\B ) = \sum_{i=0}^{N-1}\A^{i}\R(\B),
\end{equation*}
with addition corresponding to the direct sum of vector spaces and $\A^{0}$ being identity matrix.

The column space of $\B _{N\times p}$ is defined as $\mathrm{\textbf{R}}(\B )=\{\B \v |\v \in  {\R} ^p\}$.
The proof of this lemma is given as follows based on Ref. \cite{AntsaklisSLSbook}. \\
\textit{\textbf{Proof}}. 
Denoting $S = \left \{\x \Big | \x = \int_{t_0}^{t_f}\e^{\A  (t_f-s)}\B  \u(s)\mathrm{d}s, \forall \u \right\}$, from 
\begin{equation*}
\mathrm{e}^{\A} = \sum_{i=0}^{\infty}\frac{1}{i!}\A^i
\end{equation*}
we have 
\begin{eqnarray*}
S &=& \left \{\x \bigg| \x = \int_{t_0}^{t_f}  \sum_{i=0}^{\infty}\frac{1}{i!}\A^i (t_f-s)^i   \B  \u(s)\mathrm{d}s, \forall \u \right\}
\\ \nonumber &=&
\left \{\x \bigg| \x = \sum_{i=0}^{\infty} \A^i  \B \int_{t_0}^{t_f}  \frac{1}{i!} (t_f-s)^i    \u(s)\mathrm{d}s, \forall \u \right\}
\\ \nonumber &\subset &
\sum_{i=0}^{N-1}\A^{i}\R(\B) = \langle \A  | \B  \rangle.
\end{eqnarray*}
Hence we have $S \subset \langle \A  | \B  \rangle$.

Consider the matrix
$$\s_0 = \int_{t_0}^{t_f}\e^{\A  (t_f-s)}\B  \B^{\text{T}}\e^{\A^{\text{T}}  (t_f-s)} \mathrm{d}s.$$
If $\x$ belongs to the null space of $\s_0$, we have 
\begin{eqnarray*}
0 = \x^{\text{T}} \s_0 \x &=& \int_{t_0}^{t_f} \x^{\text{T}} \e^{\A  (t_f-s)}\B  \B^{\text{T}}\e^{\A^{\text{T}}  (t_f-s)} \x  \mathrm{d}s 
\\ \nonumber & =  &
\int_{t_0}^{t_f} \|  \B^{\text{T}}\e^{\A^{\text{T}}  (t_f-s)} \x \| \mathrm{d}s, 
\end{eqnarray*}
which induces $\B^{\text{T}}\e^{\A^{\text{T}}  (t_f-s)} \x = \zero$ for all $s \in [t_0, t_f]$.
This requires that all derivatives of $\B^{\text{T}}\e^{\A^{\text{T}}  (t_f-s)} \x$ equal to $\zero$ at $t_f$, that is 
$$\B^{\text{T}}\x = \zero, \B^{\text{T}}\A^{\text{T}}\x = \zero, \cdots, \B^{\text{T}}(\A^{\text{T}})^m\x = \zero, \cdots$$
which gives 
\begin{eqnarray*}
\x &\in & \N(\B^{\text{T}}) \cap  \N(\B^{\text{T}}\A^{\text{T}}) \cap \cdots \cap \N(\B^{\text{T}}(\A^{\text{T}})^m) \cap \cdots
\\ \nonumber &  &
 = [\R (\B) + \R (\A\B) + \cdots +\R (\A^{N-1}\B)]_\perp
 \\ \nonumber &  &
 = \langle \A  | \B  \rangle_\perp,
\end{eqnarray*}
since $\N(\B^{\text{T}}) = \{\x | \B^{\text{T}} \x = \zero \} = [\R (\B)]_\perp$, where $[\Q]_\perp$ means the orthogonal complementary space of $\Q$.
Conversely, if $\x \in \langle \A  | \B  \rangle_\perp$, we have $\x \in \N (\s_0)$.
Hence
$$\N (\s_0) = \langle \A  | \B  \rangle_\perp,$$
or equivalently,
$$\R (\s_0) = \langle \A  | \B  \rangle.$$
If $\x \in \langle \A  | \B  \rangle$, there exists a vector $\z$ such that $\x = \s_0 \z$.
Then using the form of the energy-optimal control signal $\u(s) = \B^{\text{T}}\e^{\A^{\text{T}}  (t_f-s)} \z $, we have 
\begin{eqnarray*}
\x =  \s_0 \z  & = &  \int_{t_0}^{t_f}\e^{\A  (t_f-s)}\B  \B^{\text{T}}\e^{\A^{\text{T}}  (t_f-s)} \z \mathrm{d}s
=
\int_{t_0}^{t_f}\e^{\A  (t_f-s)}\B  \u(s)\mathrm{d}s
 \in
 S.
\end{eqnarray*}
Thus we get $  \langle \A  | \B  \rangle \subset S$.

Taken together, we have the lemma (\ref{lemma1}):
$\left \{\x \bigg| \x = \int_{t_0}^{t_f}\e^{\A  (t_f-s)}\B  \u(s)\mathrm{d}s, \forall \u \right\}=\langle \A |\B  \rangle$.
\hspace{0.2cm}$\Box$

Based on the above lemma, we have the controllable space
\begin{equation}
\label{controllableset_main}
\Omega = \langle \A_M | \B_m \rangle + \sum_{m=1}^{M-1} \prod_{j=M}^{m+1} \e^{\A_j \tau_j}  \langle \A_m | \B_m \rangle,
\end{equation}
and reachable space  
\begin{equation}
\label{controllableset}
\sum_{m=1}^M \prod_{j=1}^m \e^{-\A _{j}\tau_j}  \langle \A _{m} | \B _{m} \rangle
= \langle \A _{1} | \B _{1} \rangle + \sum_{m=2}^M \prod_{j=1}^{m-1} \e^{-\A _{j}\tau_j}  \langle \A _{m} |  \B _{m} \rangle
\end{equation}
for a temporal network.

\subsection{Static networks}
When all snapshots of a temporal network are identical, \ie~$\A_m = \A_\textrm{s}$, the temporal network reduces to the static case \cite{Xie2002tac}, and the conditions (\ref{controllableset_main}) and (\ref{controllableset}) reduce to $\langle \A_\textrm{s} | \B \rangle$ using the fact that we can combine the product of matrix exponentials since $\A_\textrm{s}$ commutes with itself.
It follows that a static network is controllable if and only if
\begin{equation}
\label{theorm02}
\langle \A_\textrm{s} | \B \rangle = \mathbb{R}^N,
\end{equation}
which is the classic Kalman rank condition  for controllability \cite{Kalman63}.

\subsection{Relation between the controllable spaces of static vs. temporal networks}
\label{SI_RelTemSta}
\begin{figure}[H]
\centering
\includegraphics[width=1\textwidth]{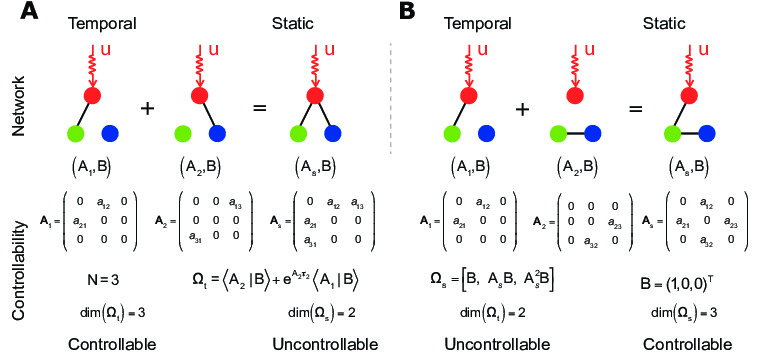}
\caption{\label{fig_notequalbetTemandAgg}
\textbf{There is no theoretically determinate relation between the controllability of temporal and static networks.}
(\textbf{A}), A controllable temporal network corresponding to an uncontrollable static network.
(\textbf{B}), A controllable static network corresponding to an uncontrollable temporal network.
Here as in the main text, we assume that one input corresponds to exactly one driver node, meaning that $\B_m$ is diagonal upon row permutation with a single entry equal to one in each column.
}
\end{figure}

After obtaining the controllable space of temporal and static networks, we ask: what is the relation between (\ref{controllableset_main}) and its counterpart (\ref{theorm02})?
This relationship between $\Omega_\text{t}$ and $\Omega_\text{s}$ is not determinate; 
Figure~\ref{fig_notequalbetTemandAgg} shows a simple contrived example illustrating that theoretically both $\Omega_\text{t} \supsetneqq\Omega_\text{s}$ 
and $\Omega_\text{t} \subsetneqq \Omega_\text{s}$ are possible.

\section{Description of empirical data sets}
\label{SI_datadescri}

We construct temporal networks from four different kinds of empirical data
(Table~\ref{fig_SI_tabel_data}) :

\textbf{ACM conference}: 
This data set is from the ACM Hypertext 2009 conference provided by the SocioPatterns collaboration \cite{datawebsite}, where the 113 conference attendees wearing radio badges were monitored for face-to-face communications.
Every communication between a pair of attendees is stored as a triplet $(t,i,j)$, meaning that people with anonymized IDs $i$ and $j$ chatted with each other during the 20-second interval from $[t-20s, t]$. 
The data spans a time period of about 2.5 days (212,340s) starting from 8am on Jun 29th, 2009 \cite{Isella2011}.
The snapshot duration $\Delta t$ is chosen from $1000$s to $212,340$s for each temporal network, yielding different number of snapshots.

\textbf{Student contacts}:
This data set consists of a sequence of contacts between students in a high school in Marseilles, France, with 126 students in three classes over 4 days in Dec. 2011 \cite{Fournet2014}.
The format of the data is same as that of ACM conference, and it is also provided by the SocioPatterns collaboration \cite{datawebsite}.
Here $\Delta t$ is chosen between $1000$s to $326,450$s.

\begin{figure}[H]
\centering
\includegraphics[width=1\textwidth]{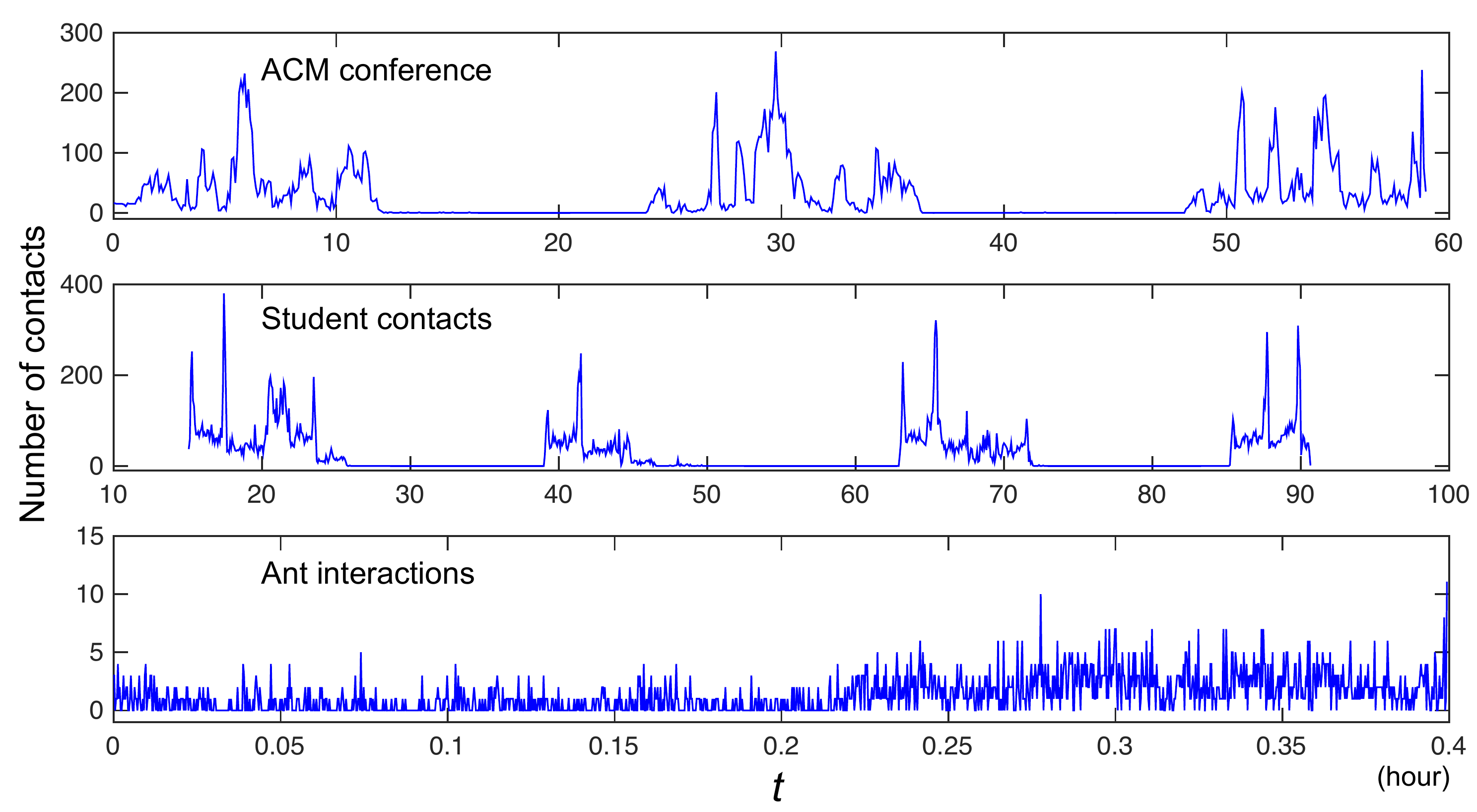}
\caption{\label{fig_SI_numberofContacts}
\textbf{Contact activity in empirical data.}
The curves show the contact activity (number of contacts over a $300$s time window) for the ACM conference and student contacts, and over 10 seconds for ant interactions.
For human interactions we observe the rhythm of day and night, while for ants the number of interactions shows little temporal variation, \emph{i.e.} with no bursts or lulls.
}
\end{figure}

\textbf{Ant interactions}: 
The \textit{Temnothorax rugatulus} interactions in this data set represent antenna-body contacts of four ant colonies \cite{BlonderAnt2011}.
We adopt the largest colony (colony $1$) as our data set, comprising 1,911 interactions between 89 ants over 1,438 seconds.
Here the duration of each antenna-body is neglected (Fig.~\ref{fig_SI_numberofContacts}).
As with the ACM conference data, we generate the snapshots of the temporal network with $\Delta t$ chosen between $10$s to $1438$s.

\textbf{Protein network}: 
This dataset is based on the protein-protein interaction network of \textit{Saccharomyces cerevisiae} from DIP (Database of Interacting Proteins), consisting of 5,023 proteins and 22,570 interactions.
From different gene expression sets, researchers construct the dynamic protein-protein interactions by identifying the active time of each protein \cite{PPIdata}. 
According to gene ontology, networks of proteins are then constructed based on three domains:
cellular component (CC), molecular function (MF), and biological process (BP).
Here we have 33, 50, and 50 snapshots for CC, MF, and BP, where the number of related proteins is 84, 74, and 85, respectively.

\textbf{Technological network}: 
For the three datasets of the technology network we considered, each contains the sequence of all data packet exchanges observed between 25 wireless radios in an emulated mobile ad-hoc network (MANET) that experiences a denial-of-service cyber-attack.  
Each dataset covers a time period of 900s, with the cyber-attack occurring after approximately 300s of operation.  
The packets are generated using real application and networking software running in a special test environment that emulates the packet loss characteristics of a wireless communication channel between radios in motion.  
A packet exchange is recorded by each network protocol that handles the packet.  
The applications generating the packets are specially designed test applications that are configured to model communication in a mobile wireless network that could be seen in a search and rescue mission.  
We construct the temporal network with 50 snapshots from three datasets named 1-ip6, 2-ip6, and 3-ip6, respectively.

Each network is on the same set of $34$ nodes.

\begin{figure}[H]
\centering
\includegraphics[width=1\textwidth]{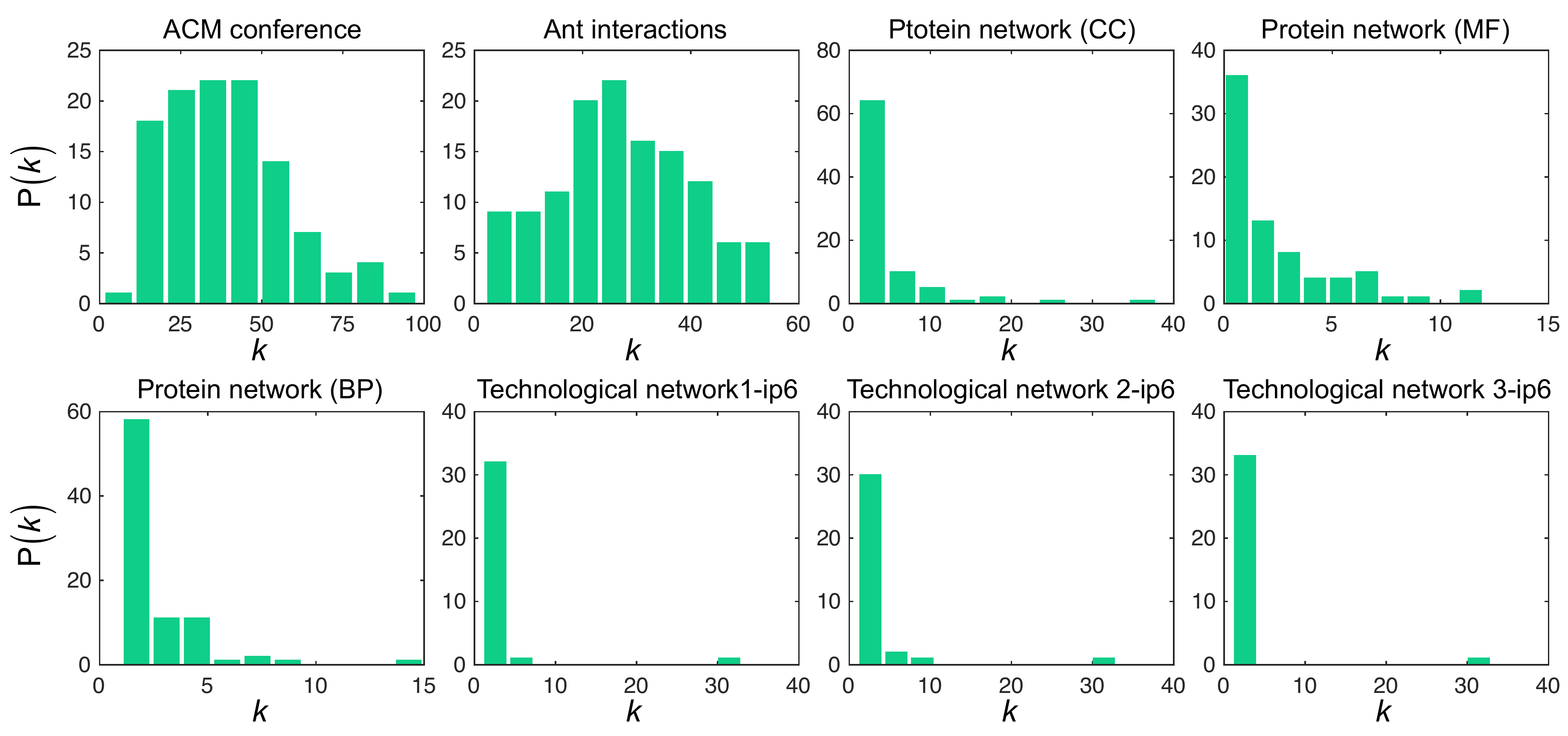}
\caption{\label{fig_SI_DegreeDis}
\textbf{Degree distribution of the static networks corresponding to four kinds of empirical datasets.}
The static networks are aggregated from all contacts for the ACM conference and ant interactions.
For protein and technological networks, the static networks are aggregated from all snapshots.
}
\end{figure}

Figures~\ref{fig_SI_numberofContacts}-\ref{fig_SI_NumComp} show, as a function of time, the interaction activity, degree distribution, average degree, and number of components of the aggregated networks for each of these datasets.

\begin{table}[H] 
\caption{
Basic information of the empirical data sets. 
$N$ is the number of nodes, while $M$ is the (maximum) number of snapshots.
For human and animal data sets stored in terms of sequence of interactions, $M$ is acquired from the time window we choose to aggregate the networks.
Considering the basic attributes of temporal networks highly depend on $\dt$, here we only list $N$ and $M$, and 
other information of the data is given in Figs.~\ref{fig_SI_numberofContacts},~\ref{fig_SI_DegreeDis},~\ref{fig_SI_AveDegree},~and~\ref{fig_SI_NumComp}.
}
\label{fig_SI_tabel_data}
\center
\begin{tabular}{cccccccccc} 
\hline \hline
  \multirow{2}*{} &   ACM  &   Student  &   Ant & \multicolumn{3}{c}{Protein network}  & \multicolumn{3}{c}{Technological network} \\ \cline{5-7}\cline{8-10}
            &conference   &contacts  &interactions   &CC &MF &BP          &1-ip6&2-ip6&3-ip6\\ \hline
$N$ & $113$ & $126$ & $89$ & 84& 74&85 &34 &34 &34  \\
$M$ & $212,340/\Delta t$ & $326,430/\Delta t$ & $1,438/\Delta t$ &33 &50 &50 &50 &50 &50  \\
\hline \hline 
\end{tabular}
\end{table}

\section{Data randomization and null models}
\label{SI_randomizdmethods}
Each of the studied data sets can be represented by a sequence of contacts,
namely triplets of the form $(t,i,j)$, meaning that individual $i$ and $j$ interact with each other at time $t$ (ant interactions) or from $t$ for a duration of 20s (ACM conference).
There are many different methods to randomize temporal networks, yielding different null models \cite{Holme2012}.
Here we focus on four typical null models:

\textbf{Time Reversal (TR)}: 
The temporal order of the contacts is reversed, \emph{e.g.} with the first becoming the last and vice versa.
Note that the duration time of each pair of individual interaction could be changed under TR, especially for high resolution data sets.
For example, if we have a sequence $(t_1,i,j)$, $(t_2,i,j)$, $(t_3,i',j')$, and $(t_3,i',j)$ (which is common in the ACM conference data set), after we implement TR, we may have $(t_3,i,j)$, $(t_2,i,j)$, $(t_1,i',j')$, and $(t_1,i',j)$ where the duration time of every link is retained, or $(t_3,i,j)$, $(t_3,i,j)$, $(t_2,i',j')$, and $(t_1,i',j)$ where the duration time of $i$ interacting with $j$ is compressed as $20$s rather than $2*20$s.
In our data, we run TR according to the two cases above, and find that results are robust.
This model is designed to assess the causality between individual interactions, for example, whether the latter contacts are triggered by the former \cite{Holme2012}, or if there are strong time-dependent patterns embedded in the original data.

\textbf{Randomly Permuted Times (RPT)}: 
Here we shuffle the timestamps of the contacts, leaving the sources and targets of the links unaltered.
Note that RPT has the effect of destroying temporal patterns and erasing time correlations between contacts.

\textbf{Randomized Edges (RE)}: 
In this model, we iteratively choose pairs of edges $(i,j)$ and $(i',j')$, and replace them with $(i, i') (j, j')$ or $(i,j') (j, i')$ with equal probability provided the change results in no self loops.
Here duration of interactions is maintained from the point of whole static network, while the numbers and durations of an individual node's contacts will be changed with high probability.
For example, for two contiguous of contacts $(t_1,i,j)$ and $(t_2,i,j)$, RE may change $(t_2,i,j)$ into $(t_2,i,j')$.
The degree of every node in the whole static network could be conserved if the data is in low resolution while will be changed in high resolution containing the example case given above.

\textbf{Randomized Edges and Randomly Permuted Times (RERPT)}: 
Equivalent to RE followed by RPT.

Since the PPI and technology networks are already represented as network snapshots, we obtain the contact sequence directly from each snapshot, where the interaction time is the order of the containing snapshot.
That is, each link (between $i$ and $j$) in snapshot $m$ is represented by the triplet $(m,i,j)$.

\begin{figure}[H]
\centering
\includegraphics[width=1\textwidth]{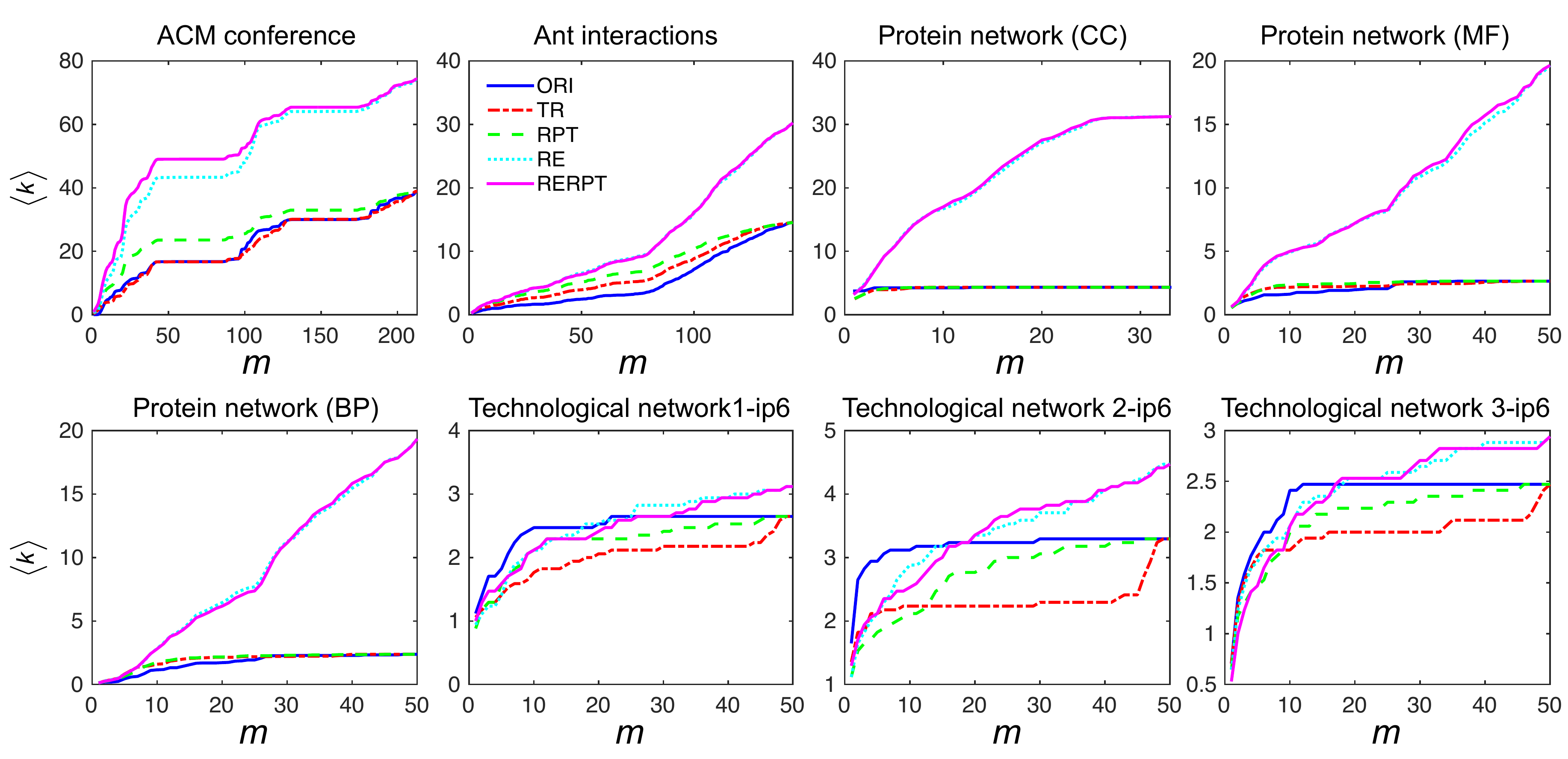}
\caption{\label{fig_SI_AveDegree}
\textbf{Average degree of static networks as a function of  snapshots aggregated.}
For the ACM conference and ant interactions network, we set $\Delta t = 1000$s and $10$s, which generates $213$ and $144$ snapshots, respectively.
For the protein network (CC) we have 33 snapshots, with 50 for each of (MF) and (BP).
``ORI'' denotes the original data sets.
The different data sets show different patterns in how nodes acquire links as the system evolves, \emph{i.e.} how $\langle k \rangle$ depends on $m$.
For the ACM conference, $\langle k \rangle$ shows long plateaus corresponding to break periods in the conference (Fig.~\ref{fig_SI_numberofContacts}).
For each network, we also show the effect of the randomization procedures discussed in Sec.~\ref{SI_randomizdmethods}:
only RE changes the ultimate value of $\langle k \rangle$;
TR and RPT change the pattern of increase of $\langle k \rangle$ but not the final value.
}
\end{figure}

\begin{figure}[H]
\centering
\includegraphics[width=1\textwidth]{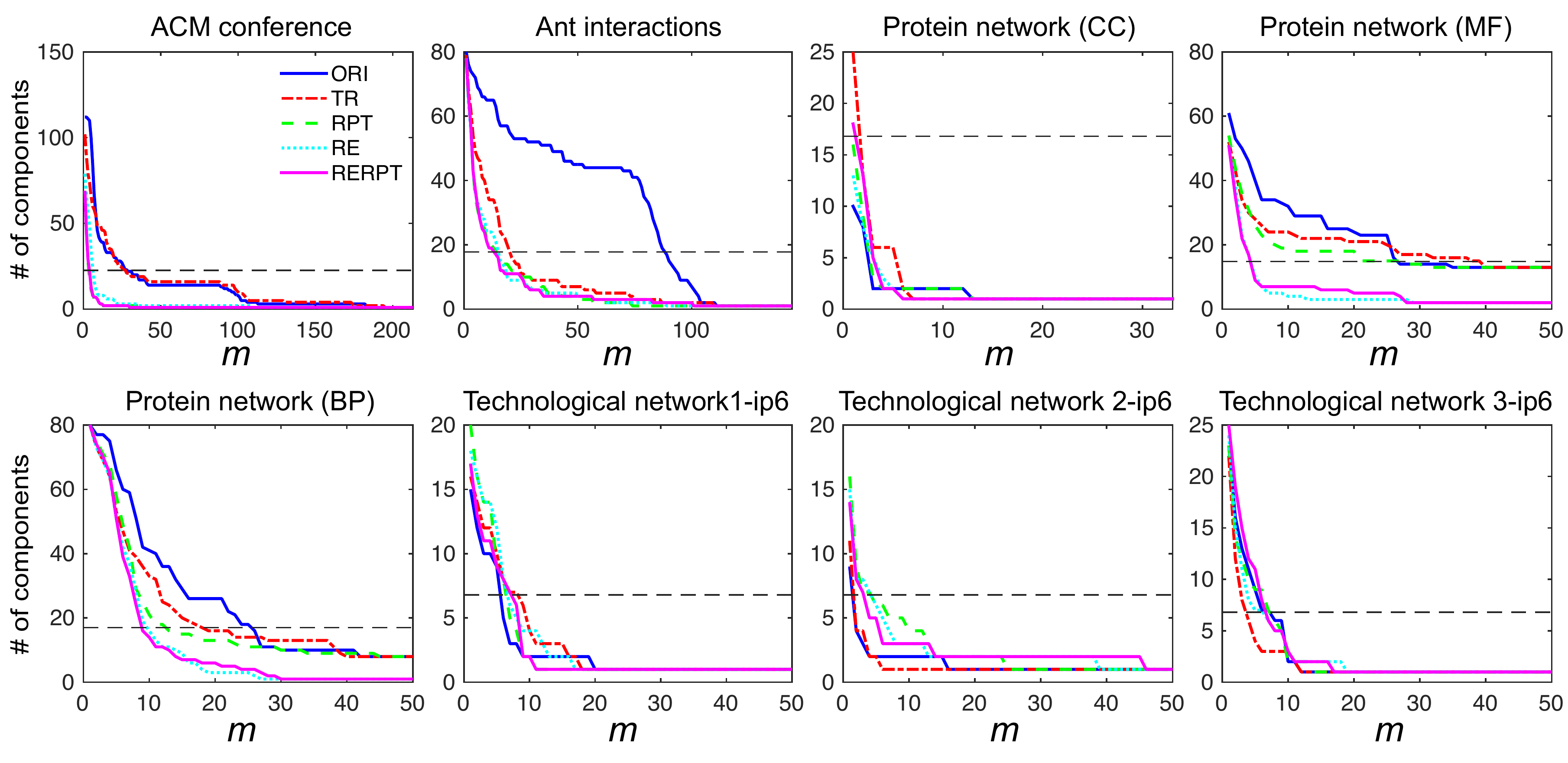}
\caption{\label{fig_SI_NumComp}
\textbf{Number of components of static networks as a function of snapshots aggregated.}
Dashed lines correspond to 20\% of the number of nodes in the network. The intersections with the colored lines therefore denote the number of snapshots that must elapse to achieve full controllability using 20\% of the network as driver nodes. 
Here, the number of components in the static equivalent can be no larger than the number of driver nodes when the temporal network is controllable.
}
\end{figure}

For the empirical data we considered, the effects of the above randomizations on the average degree and number of components of the aggregated networks are shown in Figs.~\ref{fig_SI_AveDegree} and \ref{fig_SI_NumComp}.

\section{Relationship between $S_\text{t}$ and $S_\text{s}$}
\label{relation_stss}

 For a given (sub)sequence of snapshots, we define an associated static network $\A_\textrm{s}$ by taking the element-wise average of the $\A_m$ weighted by their duration times $\tau_m$, 
$\A_\textrm{s} = \frac{1}{t_M} \sum_m \tau_m \A_m$.
This reflects the widely-used convention \cite{Slowingd13PRL} that a static network represents an aggregation of a temporal sequence, where each link weight is proportional to the fraction of time it is active.

As we have shown in the main text,
the mechanism we illustrate in Fig.~2A is the rule rather than the exception, which makes real temporal networks almost always more controllable than their static equivalents. 
Here we show additional analyses to check the robustness of our results.

As we only show temporal networks corresponding to a single $\dt$ for the ACM conference and ant interactions data in main text, more cases of $\dt$ are given in Fig.~\ref{fig_SI_antinteractions_withdurationtime}.
The result demonstrated in Fig.~\ref{fig_main_3} (namely that $S_\text{t} < S_\text{s}$) holds for other values of $\dt$.

\begin{figure}[H]
\centering
\includegraphics[width=1\textwidth]{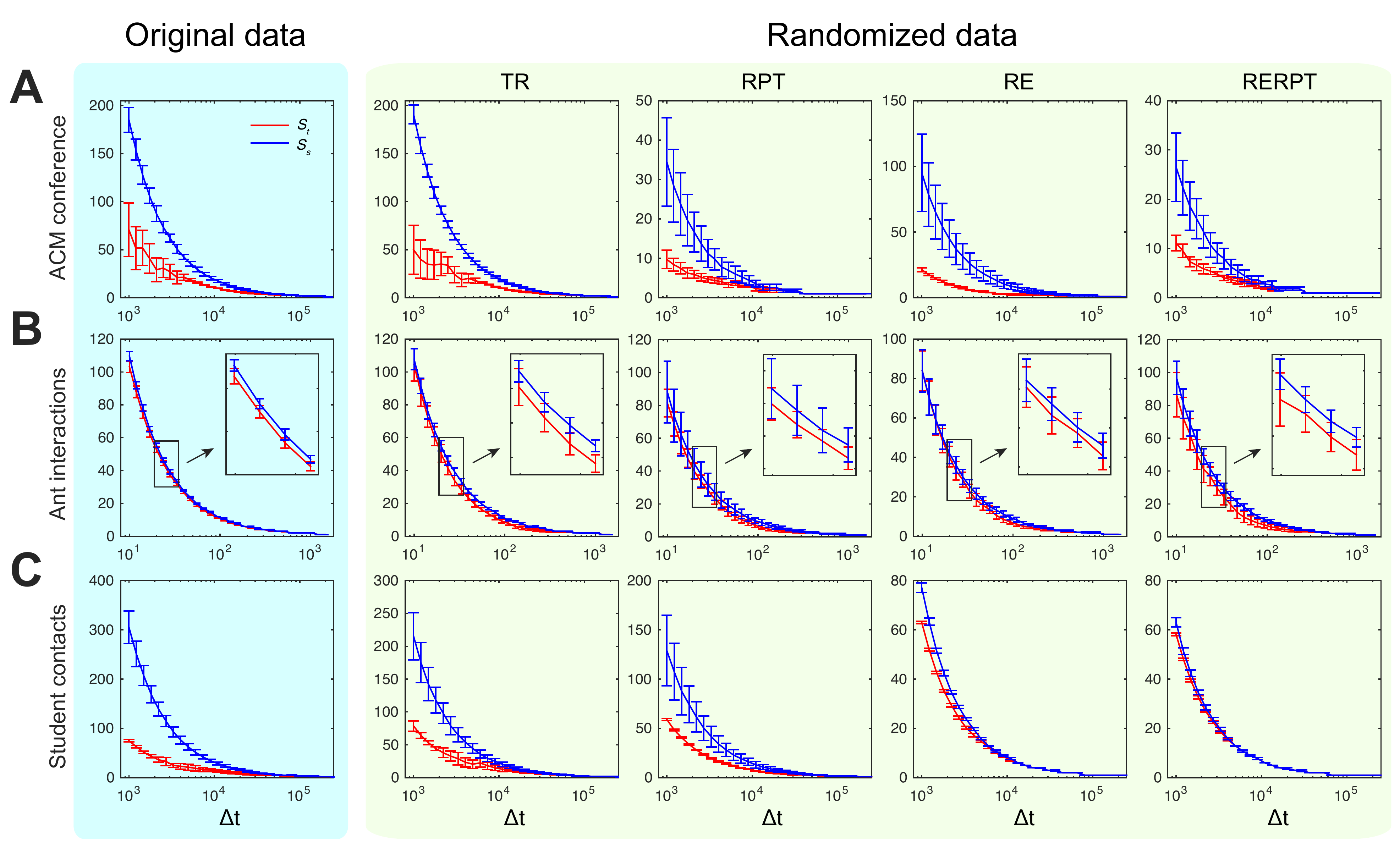}
\caption{\label{fig_SI_allDeltaT}
\textbf{Temporal networks reach controllability faster independent of the value of $\dt$.}
Shown are $S_\text{t}$ and $S_\text{s}$ for the ACM conference (\textbf{A}), ant interactions (\textbf{B}), and student contacts (\textbf{C}) networks.
Our result that temporal networks reach controllability faster holds over a wide range of $\dt$.
Parameters and other details of this analysis are the same as those used in Fig.~\ref{fig_main_2} of the main text.
}
\end{figure}

\begin{figure}[H]
\centering
\includegraphics[width=1\textwidth]{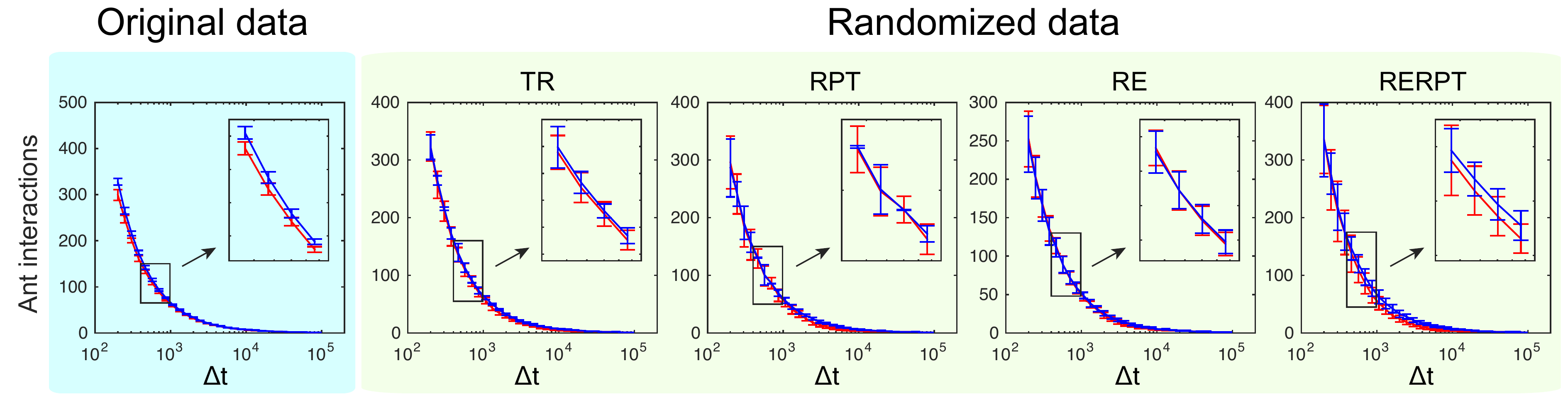}
\caption{\label{fig_SI_antinteractions_withdurationtime}
\textbf{Temporal networks reach controllability faster in the ant interaction network.}
$S_\text{t}$ is not bigger than $S_\text{s}$ for ants interactions even when each contact is equipped with a finite duration time.
Here each time point is scaled up by a factor of $60$ and every antenna-body interaction is assumed to last $20$s.
Parameters and other details of this analysis are the same as those used in Fig.~\ref{fig_main_2} of the main text.
}
\end{figure}

For the supplementary student interactions data we analyzed, the corresponding values of  $S_\text{t}$ and $S_\text{s}$ are shown in Fig.~\ref{fig_SI_allDeltaT}C.
As with the ant interactions data, the interaction times are
a sequence of discrete time points rather than a time interval like the communications between conference attendees and students.
We give the results based on the original data format in Fig.~\ref{fig_main_2}, 
assigning a small, finite duration to each contact,
which generates the result in Fig.~\ref{fig_SI_antinteractions_withdurationtime}.
The robustness of the results for other duration times has also been verified. 
The results shown in Fig.~\ref{fig_SI_antinteractions_withdurationtime} corroborate those in Fig.~\ref{fig_main_2}B,
suggesting that the result given in main text does not depend on the duration time for ants interactions.

\begin{figure}[H]
\centering
\includegraphics[width=1\textwidth]{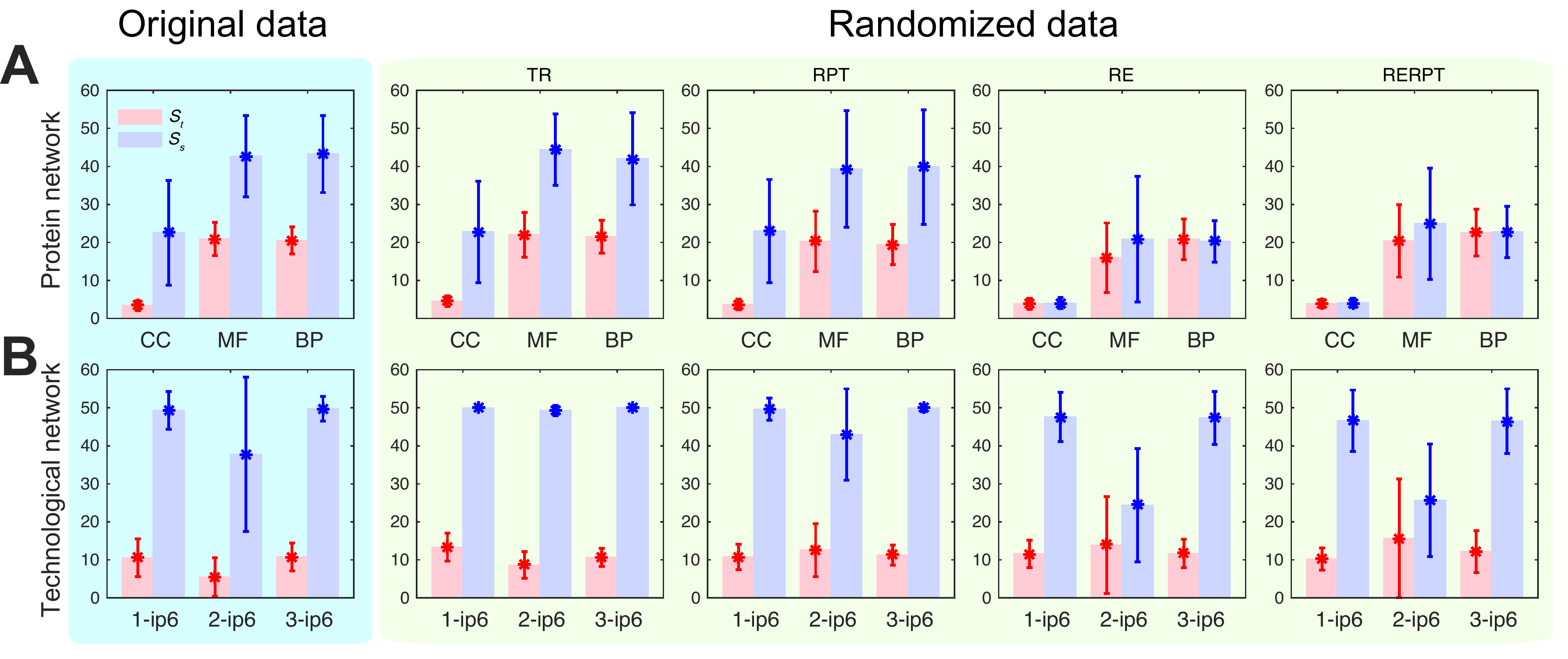}
\caption{\label{fig_SI_EightyPercentdirverNodes}
\textbf{Temporal networks reach controllability faster regardless of the number of driver nodes used.}
The static versions of the technological and protein networks sometimes remain uncontrollable at the final snapshot using sets of driver nodes corresponding to 20\% of the network, as done in Fig.~\ref{fig_main_2}.
Here we calculate $S_\text{t}$ and $S_\text{s}$ by instead using sets of driver nodes corresponding to 80\% of the network size.
Our demonstration that temporal networks reach controllability faster than their static counterparts remains true for these larger sets of driver nodes.
Predictably, both $S_\text{t}$ and $S_\text{s}$ decrease relative to Fig.~3C and 3D.
Nonetheless, we still observe cases where $S_\text{s}=M$, meaning that the static network remains uncontrollable even after the final snapshot is aggregated. 
This is true even though a full 80\% of the nodes are directly controlled. 
In contrast, the temporal version of the network \emph{is} controllable, and with only 20\% of the network as driver nodes.
Each bar corresponds to $10^3$ random sets of driver nodes.
}
\end{figure}

For a temporal network with $M$ snapshots, we define $S_\text{t}$ ($S_\text{s}$) to be equal to $M$ if the corresponding temporal (static) networks are not controllable even upon reaching (aggregating) the final snapshot $M$.
In this case, the number of snapshots required for control is larger than $M$, or equivalently, more driver nodes are needed.
For the protein and technological networks, we find many cases where $S_\text{s} = M$ in Fig.~2, hence we performed additional analysis by 
adding more driver nodes and thereby decreasing $S_\text{t}$ and $S_\text{s}$.
The results are shown in Fig.~\ref{fig_SI_EightyPercentdirverNodes}.

\section{Control energy}
\label{quadraticprogramming}
\subsection{Derivation of control energy for temporal networks}
\label{SI_defConEnergy}
For a single snapshot $(\A ,\B )$ (or equivalently, a static network), the minimum energy for controlling the system from $\x _0$ at $t_0$ to $\x_f$ at $t_f$ corresponds to the unique input of the form
$\u(t)=\B^\textrm{T}\e^{\A ^\textrm{T}(t_f-t)}\textbf{\textrm{c}}_s$, where $\textbf{\textrm{c}}_s = \W_s^{-1}(\x _f - \e^{\A t_f}\x _0)$ and
$\W_s = \int_{t_0}^{t_f}\e^{\A (t_f-s)}\B \B^{\text{T}}  \e^{\A ^{\text{T}}(t_f-s)} \mathrm{d}s$ \cite{Yan2012PRL}.
Here $\textbf{\textrm{c}}_s$ is a constant vector determined by $\x_0$, $\x_f$, $t_0$, and the system's dynamics.

According to the principle of optimality, if $\u(t)$ is the energy-optimal input to control a temporal network, then the energy accumulated over each snapshot must also be minimal for the control sub-problem of traveling between the states at the beginning and end of that snapshot.
Hence we we can write the candidate energy optimal control signals for a temporal network as
\begin{equation*}
\u(t) = \B_{m}^\textrm{T}\e^{\A ^\textrm{T}_{m}(t_{m}-t)}\textbf{\textrm{c}}_{m}
~~~\textrm{for}~t_{m-1}\leq t<t_{m},~m=1,2,\cdots,M.
\end{equation*}
Based on the above, the solution for the temporal network is
\begin{eqnarray}
\label{solutionforteme}
\x_f - \e^{\A _{M}\tau_M}\cdots \e^{\A _{1}\tau_1}\x_0  &=&
\e^{\A _{M}\tau_M} \cdots \e^{\A _{2}\tau_2} \int_{t_0}^{t_1}\e^{\A _{1}(t_1-s)}\B_{1}\u(s)\mathrm{d}s
\\ \nonumber
   & & + \cdots + \int_{t_{M-1}}^{t_M}\e^{\A _{M}(t_M-s)}\B_{M}\u(s)\mathrm{d}s
\\ \nonumber   &=&
\e^{\A _{M}\tau_M} \cdots \e^{\A _{2}\tau_2} \int_{t_0}^{t_1}\e^{\A _{1}(t_1-s)}\B_{1}\B_{1}^\textrm{T}\e^{\A ^\textrm{T}_{1}(t_{1}-s)}\mathrm{d}s \cdot \textbf{\textrm{c}}_{1}
\\ \nonumber
   & & + \cdots + \int_{t_{M-1}}^{t_M}\e^{\A _{M}(t_M-s)}\B_{M}\B_{M}^\textrm{T}\e^{\A ^\textrm{T}_{M}(t_{M}-s)}\mathrm{d}s \cdot \textbf{\textrm{c}}_{M}.
\end{eqnarray}

Using the following notations
\begin{eqnarray}
\nonumber
\textbf{\textrm{d}}
& = &
\x_f - \e^{\A _{M}\tau_M}\cdots \e^{\A _{1}\tau_1}\x_0,
\\ \nonumber
\W_{m}
& = &
\int_{t_{m-1}}^{t_m}\e^{\A _{j}(t_m-s)}\B_{m}\B_{m}^\textrm{T}
\e^{\A ^\textrm{T}_{m}(t_{m}-s)}\mathrm{d}s
=
\W_{m}[t_{m-1}, t_m] 
\\ \nonumber
& = &
\int_{0}^{\tau_m}\e^{\A _{m} \tau }\B_{m}\B_{m}^\textrm{T}
\e^{\A ^\textrm{T}_{m} \tau }\mathrm{d} \tau
=
\W_{m}[0, \tau_m],
\\ \nonumber
\textbf{\textrm{c}}
& = &
 \left(\textbf{\textrm{c}}^{\textrm{T}}_{1},\textbf{\textrm{c}}^{\textrm{T}}_{2},\cdots,\textrm{\textbf{c}}^{\textrm{T}}_{M} \right) ^ {\textrm{T}},
\\ \nonumber
\textbf{\textrm{H}}
& = &
\left( \e^{\A _{M}\tau_M} \cdots \e^{\A _{2}\tau_2} \W_{1}, \cdots,\e^{\A _{M}\tau_M} \cdots \e^{\A _{{m+1}}\tau_{m+1}} \W_{m} ,\cdots,\W_{M} \right)
 =
\s \W,
\\
\s & = &
 \left(\e^{\A _{M}\tau_M} \cdots \e^{\A _{2}\tau_2} , \cdots,\e^{\A _{M}\tau_M} \cdots \e^{\A _{{m+1}}\tau_{m+1}} ,\cdots,\textrm{\textbf{I}}_N \right)
\\ \nonumber
 & = &
 \left(  \prod_{l=M}^2 \e^{\A _{l}\tau_l} , \cdots,  \prod_{l=M}^{m+1} \e^{\A _{l}\tau_l}  ,\cdots,\textrm{\textbf{I}}_N \right),
 \label{es}
 \\ \nonumber
 \W & = & \text{diag}(\W_{1},\W_{2},\cdots,\W_{M}),
\end{eqnarray}
we can write (\ref{solutionforteme}) as $\textbf{\textrm{d}} = \textrm{\textbf{Hc}}$.

The energy to control temporal networks from $\x _0$ at $t_0$ to $\x _f$ at $t_f$ can be written as
\begin{equation*}
\label{energydef}
\E(\x _0, \x _f)
=
\frac{1}{2}\int_{t_0}^{t_f}\u^{\textrm{T}}(t)\u(t)\textrm{d}t
  =
  \frac{1}{2}\textrm{\textbf{c}}^\textrm{T}\W\textbf{\textrm{c}}.
\end{equation*}
Hence, the minimum energy could be obtained by solving the quadratic programming problem
\begin{eqnarray}
\label{quadraticprogrammingsi}
  \textrm{min} & ~~~~~& \E (\x _0, \x _f)
  =
  \frac{1}{2}\textrm{\textbf{c}}^\textrm{T}\W\textbf{\textrm{c}} \nonumber \\
  \textrm{s.t.} & & \textrm{\textbf{Hc}}=\textbf{\textrm{d}}
\end{eqnarray}
for the unknown $\textrm{\textbf{c}}$.

\subsection{Solving the quadratic problem}
\label{SolvingQuadradicProblem}
Since $\W$ is symmetric, we have $\W = \U \Lambda \U ^\textrm{T}$, where $\U\U^\textrm{T}=\U^\textrm{T}\U=\textrm{\textbf{I}}$ and $\Lambda$ is diagonal.
Using $$\x  = \sqrt{\Lambda} \U ^\textrm{T} \textrm{\textbf{c}}$$
 (\emph{i.e.} $\textrm{\textbf{c}} = \U \left(\sqrt{\Lambda}\right)^{-1}\x $) and
$$\K = \h \U \left(\sqrt{\Lambda}\right)^{-1}$$ 
we transfer the quadratic programming problem (\ref{quadraticprogrammingsi}) to
\begin{eqnarray}
\label{basic2}
  \textrm{min} &~~~& \E(\x_0, \x_f)=\frac{1}{2}\textrm{\textbf{c}}^\textrm{T}\W\textrm{\textbf{c}} =\frac{1}{2}\textrm{\textbf{c}}^\textrm{T} \U \Lambda \U ^\textrm{T} \textrm{\textbf{c}} = \frac{1}{2}\x  ^\textrm{T} \x 
  \nonumber \\
  \textrm{s.t.} & & \h \U \left(\sqrt{\Lambda}\right)^{-1}\x = \K\x  =\textbf{\textrm{d}}.
\end{eqnarray}

To solve (\ref{basic2}), let
\begin{equation*}
f( \x , \textrm{\textbf{v}} ) = \frac{1}{2}\x ^\textrm{T}\x + \textrm{\textbf{v}} ^ \mathrm{T} \left( \textrm{\textbf{Kx}} - \textbf{\textrm{d}} \right)
\end{equation*}
and minimize $f \left( \x , \textrm{\textbf{v}} \right)$, where $\textrm{\textbf{v}} $ is a set of Lagrange multipliers.
At the point $\left( \x ^*, \textrm{\textbf{v}}^* \right)$ where  $f \left( \x , \textrm{\textbf{v}} \right)$ reaches the minimum, the following relations must be satisfied
\begin{eqnarray}
\label{f1}
\frac{ \partial f \left( \x , \textrm{\textbf{v}} \right) }{\partial \x ^* }
 &=&
\x^* + \textbf{\textrm{K}}^\mathrm{T} \textrm{\textbf{v}}  ^ * = \zero
\\
\frac{\partial f \left( \x , \textrm{\textbf{v}} \right) }{\partial \textrm{\textbf{v}}^*}
&=&
\K\x ^* - \d = \zero.
\label{f2}
\end{eqnarray}
Multiplying both sides of (\ref{f1}) by $\K$ on the left, we have
\begin{eqnarray*}
\K\x^* + \K \textbf{\textrm{K}}^\mathrm{T} \textrm{\textbf{v}}^ *  &=& \zero
\\ 
\K\x^*  &=&  \d .
\end{eqnarray*}
If $\K \textbf{\textrm{K}}^\mathrm{T}$ is non-singular, $\textrm{\textbf{v}}  ^ * = -\left( \K \textbf{\textrm{K}}^\mathrm{T} \right)^{-1}\d$,
and then according to (\ref{f1}) we have
\begin{equation}
\label{xstar}
\x   ^ * =\textbf{\textrm{K}}^\mathrm{T} \left( \K \textbf{\textrm{K}}^\mathrm{T} \right)^{-1}\d.
\end{equation}

Since  $\K = \h \U \left(\sqrt{\Lambda}\right)^{-1}$, if we prove $\K \textbf{\textrm{K}}^\mathrm{T}$ is non-singular, then the problem (\ref{basic2}) can be solved according to the expression (\ref{xstar}).

\vspace{0.5cm}

\textit{\textbf{Proposition}}: If $\K$ is a matrix over real numbers with size $n \times m$, then the rank of $\K$ and $\K\K^\textrm{T}$ is equal.

\textbf{Proof:}
The null space of $\K^\textrm{T}$  is given by vectors $\x $ satisfying $\K^\textrm{T}\x =\textrm{\textbf{0}}$.
And the null space of $\K\K^\textrm{T}$  is given by vectors $\textrm{\textbf{y}}$ satisfying $\K\K^\textrm{T}\textrm{\textbf{y}}=\textrm{\textbf{0}}$.
Since $\K^\textrm{T}\x =\textrm{\textbf{0}}$, we have $\K\K^\textrm{T}\x =\textrm{\textbf{0}}$, \emph{i.e.} $\x $ belongs to  the null space of $\K\K^\textrm{T}$.
From $\K\K^\textrm{T}\textrm{\textbf{y}}=\textrm{\textbf{0}}$, we have $\textrm{\textbf{y}}^\textrm{T}\K\K^\textrm{T}\textrm{\textbf{y}}=\textrm{\textbf{0}}=(\K^\textrm{T}\textrm{\textbf{y}})^\textrm{T}\K^\textrm{T}\textrm{\textbf{y}}$, \emph{i.e.}, $\K^\textrm{T}\textrm{\textbf{y}} = \textrm{\textbf{0}}$ and $\textrm{\textbf{y}}$ belongs to the null space of $\K^\textrm{T}$.
Thus, the two equations $\K^\textrm{T}\x =\textrm{\textbf{0}}$ and $\K\K^\textrm{T}\textrm{\textbf{y}}=\textrm{\textbf{0}}$ have same solutions.
As such,
the number of independent vectors in the fundamental system is also the same, \emph{i.e.}
$n- \textrm{rank}(\K^\textrm{T}) = n - \textrm{rank}(\K\K^\textrm{T})$.
Hence we have rank$(\K)$ =rank$(\K^\textrm{T})$ =rank$(\K\K^\textrm{T})$. 
\hspace{10cm}$\Box$

Based on the above \textit{\textbf{Proposition}}, we have rank$(\K\K^\textrm{T})$ = rank$(\K)$ = rank$(\h \U \left(\sqrt{\Lambda}\right)^{-1})$, and $\K$ is a matrix with size $N \times NM$,
and
\begin{eqnarray*}
\h \U \left(\sqrt{\Lambda}\right)^{-1}
& = &\s \W\U \left(\sqrt{\Lambda}\right)^{-1}
= \s\U \Lambda \U ^\textrm{T} \U \left(\sqrt{\Lambda}\right)^{-1}
= \s\U   \sqrt{\Lambda}
\\ \nonumber
& = &
 \left(\mathrm{e}^{\A _{M}\tau_M} \cdots \mathrm{e}^{\A _{2}\tau_2} , \cdots, \mathrm{e}^{\A _{M}\tau_M} \cdots \mathrm{e}^{\A _{{j+1}}\tau_{j+1}} ,\cdots,\textrm{\textbf{I}}_N \right)
 \U   \sqrt{\Lambda}.
\end{eqnarray*}
We know that $\U = (u_1,u_2,\cdots,u_N)$, where $u_i$ is an eigenvector of one of the eigenvalues of $\W$, and rank$(\U)=N$.
In addition, we have rank$(\sqrt{\Lambda}) = N$.
Thus we obtain rank$(\s\U\sqrt{\Lambda})$ = rank$(\s\U)$ = rank$(\s)$.
Since the last block of $\s$ is $\textrm{\textbf{I}}_N$, we have rank$(\s)=N$.
Hence we have
rank$(\K\K^\textrm{T})$ = rank$(\K)$ = rank$\left(\h \U \left(\sqrt{\Lambda}\right)^{-1}\right)= N$, and 
$\K\K^\textrm{T}$ is a non-singular square matrix with size $N$.

Thus the solution of the problem (\ref{basic2}) is $\x ^* = \K^\textrm{T} (\K\K^\textrm{T})^{-1} \textbf{\textrm{d}}$, where $\K = \h \U \left(\sqrt{\Lambda}\right)^{-1}$.
Hence we have
\begin{eqnarray*}
\E(\x_0, \x_f) 
&=& \frac{1}{2}\x  ^{*\textrm{T}} \x ^*
=
 \frac{1}{2} \d^\textrm{T} \left[\K^\textrm{T} \left(\K\K^\textrm{T}\right)^{-1} \right] ^{\textrm{T} }  \left[\K^\textrm{T} \left(\K\K^\textrm{T}\right)^{-1} \right]  \d
\nonumber \\
& = & \frac{1}{2} \d^\textrm{T} \left(\K\K^\textrm{T}\right)^{-1} \d
=
 \frac{1}{2} \d^\textrm{T}  \left( \s\U \Lambda \U^\textrm{T}\U \Lambda ^{-1} \U^\textrm{T}  \U \Lambda \U ^\textrm{T} \s^\textrm{T} \right) ^ {-1} \d     
\nonumber \\
& = & \frac{1}{2} \d^\textrm{T}  \left( \s\W\s^\textrm{T} \right) ^ {-1} \d.
\end{eqnarray*}

\subsection{Minimum energy needed to control temporal networks}
\label{SI_MinTemCon}
Taken together, the quadratic programming problem given in (\ref{quadraticprogrammingsi}) is solved analytically by the optimal solution 
\begin{equation}
\label{solutionc}
\textrm{\textbf{c}}^* = \s^\mathrm{T} \left( \s\W\s^\mathrm{T}  \right)^{-1}\d, 
\end{equation}
with the corresponding
minimum control energy
\begin{equation}
\label{retempsi}
\E^*(\x_0, \x_f)  =  \frac{1}{2} \d^T \W_{\textrm{eff}} ^{-1} \d,
\end{equation}
where the $N \times N$ matrix $\Weff =  \s\W\s^\mathrm{T} $ is an ``effective'' gramian matrix, encoding the energy structure of the temporal network.  
Hereafter, we refer to the minimum control energy $\E^*(\x_0, \x_f)$ as simply the \emph{control energy} $\E$. 

For controllability in the case $\x_0 = \zero$, above results reduce to 
\begin{equation*}
\textrm{\textbf{c}}^*_c = \s^\mathrm{T} \left( \s\W\s^\mathrm{T}  \right)^{-1}\x_f,
\end{equation*}
and 
\begin{equation}
\label{MiniEtemp}
\E =  \frac{1}{2} \x_f^T \W_{\textrm{eff}} ^{-1} \x_f.
\end{equation}

\subsection{Minimum energy needed to control static networks}
\label{SI_tempDegSta}

When all snapshots are identical 
($\A_m = \A _{s}$), 
our results reduce to the case for static networks.
Indeed, for static networks, the quadratic programming (\ref{quadraticprogrammingsi}) becomes 
\begin{eqnarray*}
\label{quadraticprogrammingsi}
  \textrm{min} &~~~& \E (\x _0, \x _f)
  =
  \frac{1}{2}\textrm{\textbf{c}}^\textrm{T}\W\textbf{\textrm{c}} \nonumber \\
  \textrm{s.t.} & & \W_{s}\textbf{\textrm{c}}=\textbf{\textrm{d}}=\x_f - \e^{\A _{s}t_f}\x_0.
\end{eqnarray*}
If the system is controllable, $\W_{s}$ is nonsingular, and so there is a unique solution 
$\textbf{\textrm{c}}=\W_{s}^{-1} \textbf{\textrm{d}}$.
Hence, the optimal solution from above quadratic programming determines the optimal input as 
\begin{equation*}
\u(t)=\B^{\textrm{T}}\e^{\A ^\textrm{T}(t_{f}-t)}\W_{s}^{-1}\left(\x_f - \e^{\A _{s}t_f}\x_0\right),
\end{equation*}
by which we have
\begin{equation}
\label{restatic}
\E =\left(\x_f - \e^{\A _{s}t_f}\x_0\right)^{\textrm{T}} \W_{s}^{-1} \left(\x_f - \e^{\A _{s}t_f}\x_0\right),
\end{equation}
where
$\W_{s}=\int_{t_{0}}^{t_f}\e^{\A _{s}(t_f-s)}\B_{s}\B_{s}^\textrm{T}\e^{\A _{s}^\textrm{T}(t_{f}-s)}\mathrm{d}s$.
The result for $M=1$ is same as that given in \cite{Yan2012PRL}.

As $\A _i = \A $ for $i=1,2,\cdots,M$, we have
\begin{eqnarray*}
\label{temptostat}
\textbf{\textrm{SWS}}^\textrm{T}
& = &
\sum_{i=1}^{M-1}
\left(
\prod_{k=M}^{i+1}
\e^{\A _{k}\tau_k}
\W_i
\prod_{l=i+1}^{M} \e^{\A ^\textrm{T}_{l}\tau_l}
\right)
+\W_M
\nonumber \\
&= &
\sum_{i=1}^{M-1}
\int_{0}^{\tau_i}\e^{\A  (\tau+\sum_{k=i+1}^M \tau_k) }\B\B^\textrm{T}
\e^{\A ^\textrm{T} (\tau+\sum_{k=i+1}^M \tau_k) }\mathrm{d} \tau
+\W_M
\nonumber \\
&= &
\int_{0}^{\sum_{k=1}^M \tau_k}\e^{\A  \tau }\B\B^\textrm{T}\e^{\A ^\textrm{T} \tau }\mathrm{d} \tau
\nonumber \\
&= &
\W_{s}.
\end{eqnarray*}
Thus the energy for controlling temporal networks (\ref{MiniEtemp}) reduces to the result for static networks (\ref{restatic}).

\section{Analysis of the control energy}
\label{SI_En_numberical}

To account for the fact that the control energy generally grows as $\x_0$ and $\x_f$ become further apart, we can write the \emph{normalized control energy} \cite{Yan2012PRL}
from $\x_0=\textbf{0}$ to $\x_f$ as
\begin{equation*}
\E
 =  \frac{\textrm{\textbf{x}}_f^\textrm{T}  \left( \textrm{\textbf{S}}\textrm{\textbf{W}}\textrm{\textbf{S}}^\textrm{T} \right) ^ {-1} \textrm{\textbf{x}}_f }{ 2\textrm{\textbf{x}}_f^\textrm{T}\textrm{\textbf{x}}_f}.
\end{equation*}
Irrespective of $\textrm{\textbf{x}}_f$, we can obtain bounds of $\E$ for every $\| \textrm{\textbf{x}}_f \| = 1$ are
\begin{equation*}
\Elb = \frac{1}{2\lambda_{\text{max}}} \leq  \E \leq  \Eub = \frac{1}{2\lambda_{\text{min}}},
\end{equation*}
where $\Elb$ and $\Eub$ are the lower and upper bound of $E$.
$\lambda_{\text{max}}$ and $\lambda_{\text{min}}$  are the maximum and minimum eigenvalues of $\W_{\textrm{eff}}$, respectively.

\begin{figure}[H]
\centering
\includegraphics[width=1\textwidth]{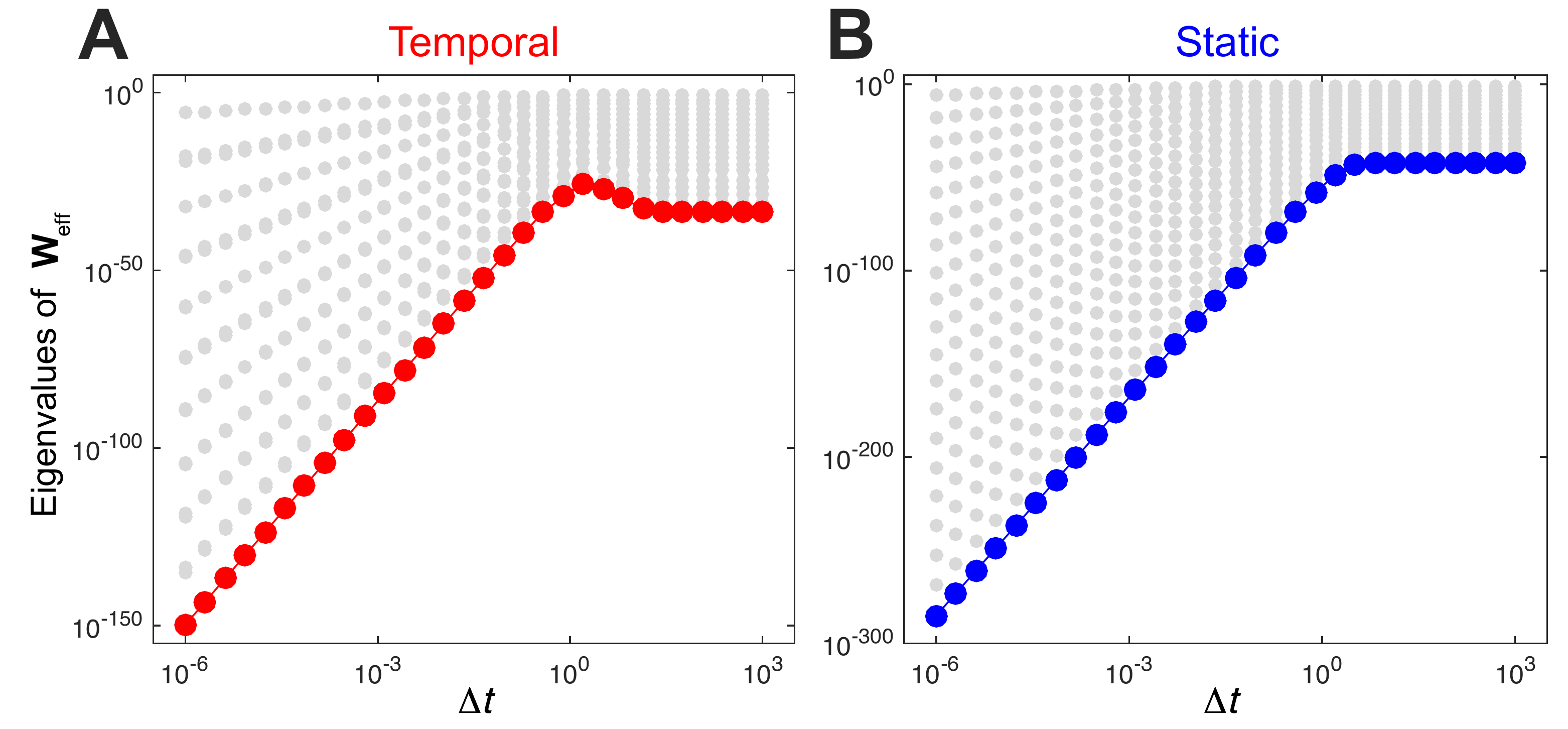}
\caption{\label{SI_eig}
\textbf{The minimum eigenvalue of $\textbf{\textrm{W}}_\text{eff}$ dominates the control energy.}
For different $\Delta t$, all eigenvalues of $\textbf{\textrm{W}}_\text{eff}$ are given by gray points for (\textbf{A}) temporal and (\textbf{B}) static networks, where the minimum eigenvalues are enlarged in red and blue, respectively.
The eigenvalues of $\textbf{\textrm{W}}_\text{eff}$ vary over many orders of magnitude, implying that the average control energy is dominated by the worst-case direction (corresponding to the $\lambda_\text{min}$) of $\textbf{\textrm{W}}_\text{eff}$ (eq.~(\ref{retempsi})).
Since $\lambda_\text{min}$ is much greater for 
the temporal network, the energy required to move in typical control directions is thus expected to be less than in the
corresponding static network.
Here the system parameters are the same as those used in Fig.~\ref{fig_main_3}A of the main text.  
}
\end{figure}

We can demonstrate numerically that $\eig$ is generally greater in temporal networks, and hence $\Eub$ is usually smaller, often much smaller, than in their static equivalents (Fig.~\ref{SI_eig}).  
This implies that the average control energy $\Eavg$ is typically much less in a temporal network, despite the fact that $\Eub$ may correspond to different ``worst-case'' directions in the static vs. temporal case. 
Indeed a typical control direction $\d$ will have some component lying along the eigenvector of $\Weff$ corresponding to $\eig$. 
Since the eigenvalues of $\Weff$ typically vary many orders of magnitude (Fig.~\ref{SI_eig}),
this worst-case direction dominates the control energy, and $\Eub$ is expected to be representative of $\Eavg$, an expectation borne out by our results (Fig.~\ref{fig_main_3}). 
This also explains why the scaling of $\Eavg$ is determined by that of $\Eub$, which we can show decreases according to $\Eavg \sim \dt ^ {- \gamma}$ for small $\dt$ before reaching a plateau (Fig.~\ref{fig_main_3}). 

\begin{figure}[H]
\centering
\includegraphics[width=1\textwidth]{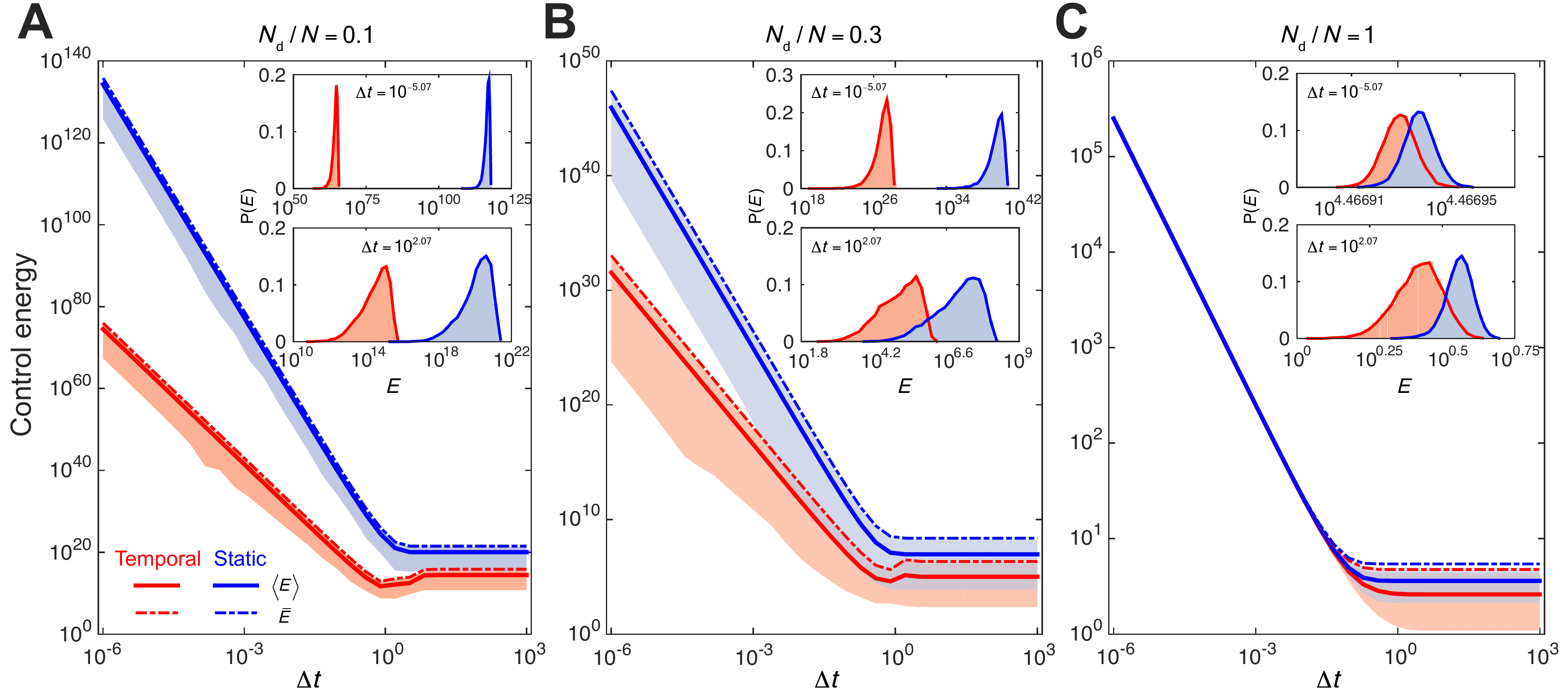}
\caption{\label{fig_SI_energyN10M2k4_minus3_minus1}
\textbf{Temporal networks require less control energy compared to static networks.}
Counterpart to Fig.~\ref{fig_main_3} of the main text
with $N=10, M=2, \bar{k}=4$, $a_1 = -3$, and $a_2 = -1$.
}
\end{figure}

\begin{figure}[H]
\centering
\includegraphics[width=1\textwidth]{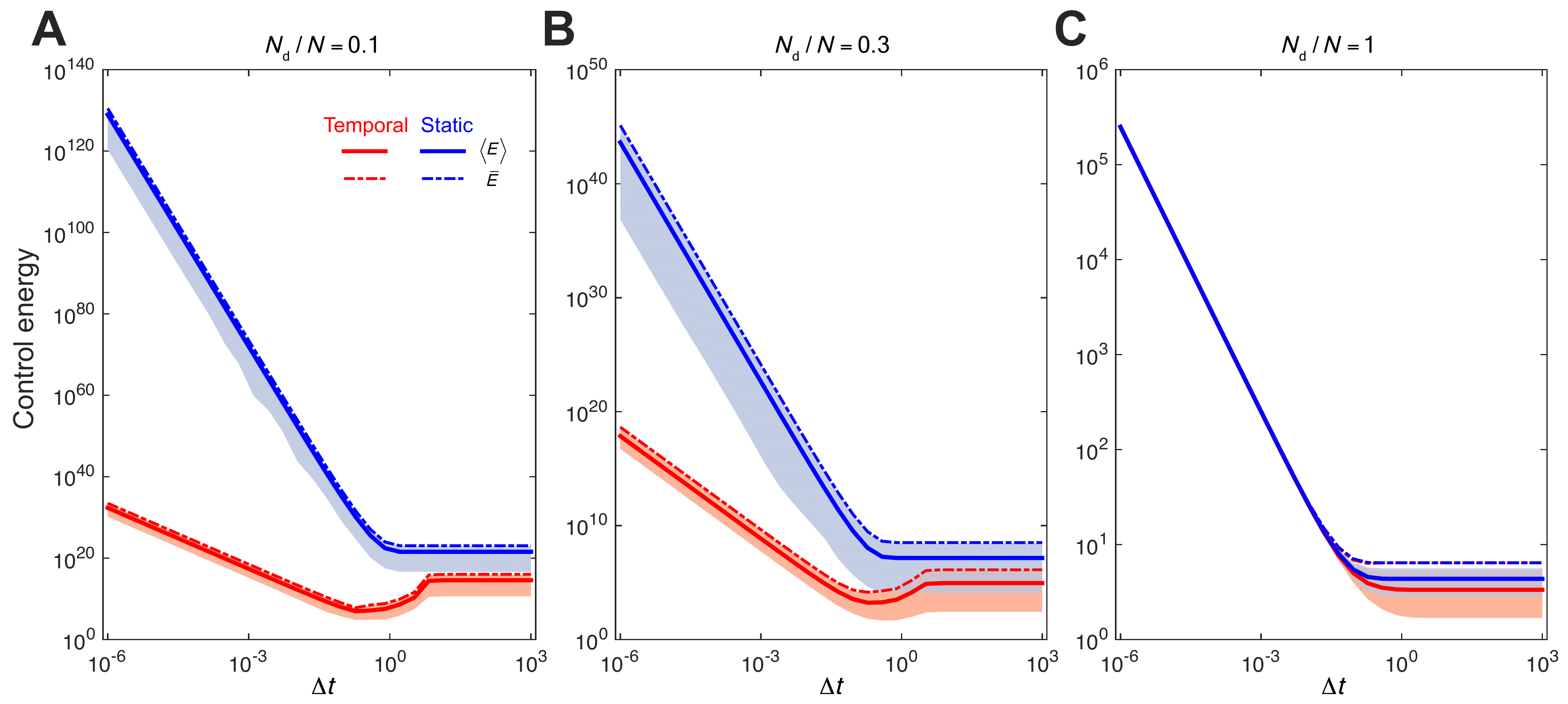}
\caption{\label{fig_SI_energyN10M5k4minus2all}
\textbf{Temporal networks require less control energy compared to static networks .}
Counterpart to Fig.~\ref{fig_main_3} of the main text
with $N=10, M=5, \bar{k}=4$, and $a_i = -2$ for $i=1,2,\cdots,5$.
}
\end{figure}

\begin{figure}[H]
\centering
\includegraphics[width=1\textwidth]{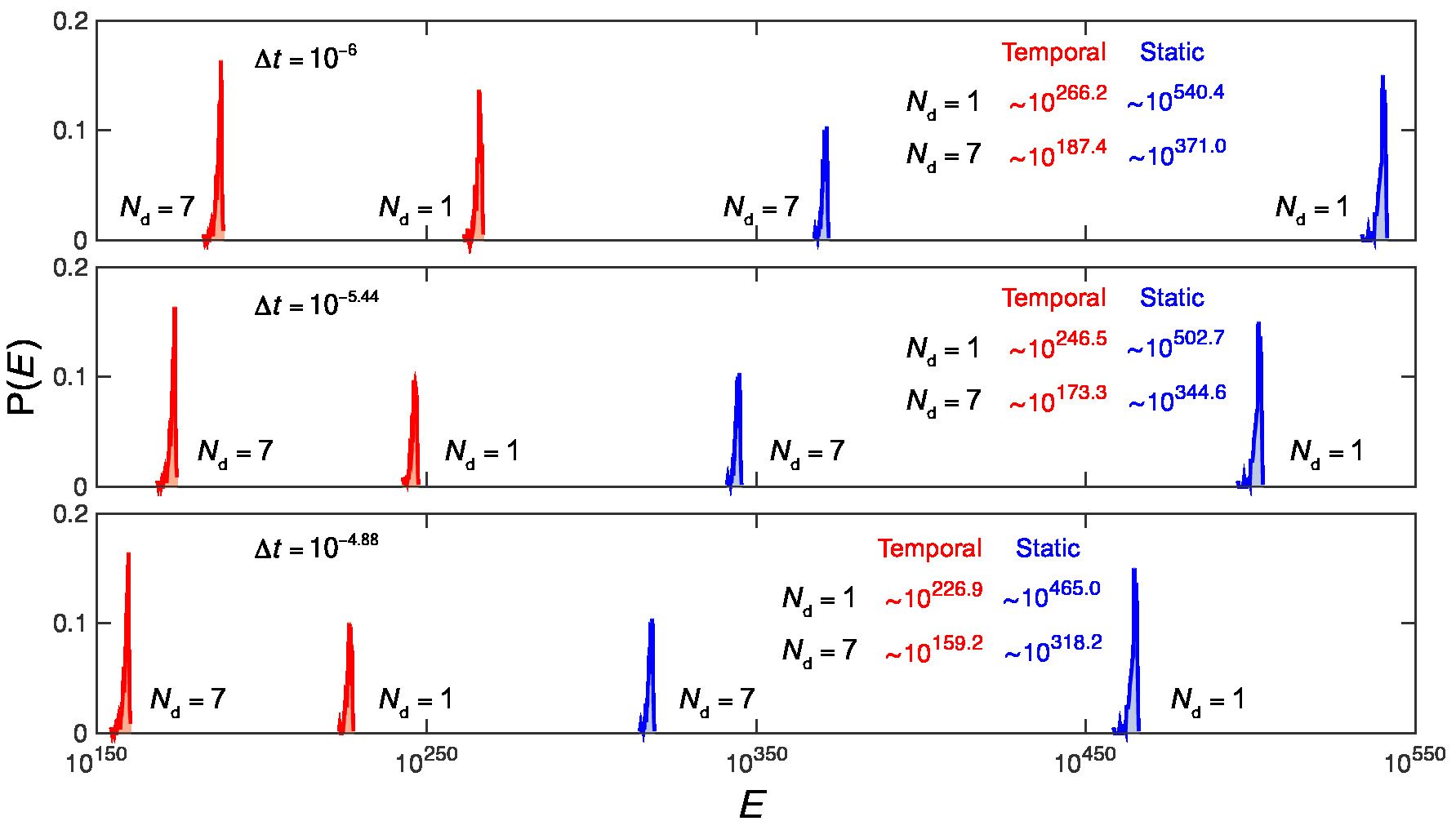}
\caption{\label{fig_SI_energy_N34Ndmore}
\textbf{Difference in control energy for a real network.}
We aggregate the total $M=50$ snapshots of the 1-ip6 network into $M=2$ snapshots
We show the distribution of the control energy over $300$ randomly-selected final states with unit distance away from $\x_0 = \zero$, for varying $\dt$ and numbers of driver nodes $N_d$ (blue: static, red: temporal).
The corresponding average energies $\Eavg$ are denoted in each panel.
We find that control energy decreases as either $N_d$ or $\dt$ increases.
Here we choose $a_1 = -1$ and $a_2=-2$.
}
\end{figure}

The robustness of these results has been checked for other networks and 
shown
in Figs.~\ref{fig_SI_energyN10M2k4_minus3_minus1}
-\ref{fig_SI_energy_N34Ndmore}.

\section{Use of the Laplacian matrix for $\A _m$}
\label{LaplacianMatrix}

To study the control energy and locality,
we also creat synthetic temporal networks by generating each snapshot randomly and independently according the Erd\H{o}s-R\'enyi $G(n,p)$ model. 
We discard and re-generate any snapshot that is disconnected and assign link weights independently and randomly from $(0,1)$. 
For the control energy we add self-loops with identical weight $a_m$ to all nodes, where $a_m$ 
is chosen large enough to stabilize the standalone dynamics of each snapshot $m$, reflecting the fact that most real systems have a stable state \cite{RMay}, and its values do not change our findings.
Note, however, that our theory also works for unstable dynamics.

For the control energy and locality analyses of the main text, we employ the Laplacian matrix with self loops for the system matrix $\A_m$ of each snapshot.
Specifically, $\textbf{\textrm{L}} = (l_{ij})_{NN}$, where
\begin{eqnarray*}
l_{ij}
&=&
\begin{cases}
w_{ij}
& \text{$i \neq j$}  \\
-\sum_{j=1, j \neq i}^{N} w_{ij}
& \text{$i=j$}
 \end{cases}
\end{eqnarray*}
and $w_{ij}$ is (randomly-chosen) weight of the edge from node $j$ to node $i$.
For an arbitrary vector $\xi = (\xi_1, \xi_2, \cdots, \xi_N)^\textrm{T}$, we have
\begin{eqnarray*}
\xi^\textrm{T} \textbf{\textrm{L}} \xi &=& \sum_{i=1}^N \sum_{j=1}l_{ij}\xi_i \xi_j
=\sum_{i=1}^N \sum_{j=1, j \neq i}^N \left(w_{ij} \xi_i \xi_j - w_{ij} \xi_i^2\right)
=
-\frac{1}{2}\sum_{i=1}^N \sum_{j=1}^N w_{ij}  \left( \xi_i - \xi_j   \right)^2
 \end{eqnarray*}
Thus we know when all $w_{ij} > 0~(w_{ij} < 0)$, $\textrm{\textbf{L}}$ is negative (positive) semi-definite.
Here, we let $\A _m = a_m\textrm{\textbf{I}} + \textrm{\textbf{L}}$, where $a_m$ is chosen to stabilize the dynamics of each individual snapshot:
\begin{enumerate}
\item[*] When $w_{ij}>0$, we can tune $a_m$ to make $\A _m$ negative ($a_m<0$), negative semi- ($a_m=0$), and non-negative definite ($a_m>0$, here when $a_m$ is sufficiently positive, $\A _m$ can be positive definite) matrix $\A _m$ since $a_m$ is the maximum eigenvalue of $\A _m$.
\item[*] When $w_{ij}<0$, we can tune $a_m$ to make $\A _m$ positive ($a_m>0$), positive semi- ($a_m=0$), and non-positive definite ($a_m<0$, here when $a_m$ is sufficiently negative, $\A _m$ can be negative definite) matrix $\A_m$ since $a_m$ is the minimum eigenvalue of $\A _m$.
\end{enumerate}

\section{Locality of the optimal control trajectories for temporal networks}
\label{SI_contraj}

For the optimal solution $\textrm{\textbf{c}}^*$ shown in (\ref{solutionc}), 
we calculate the minimum control energy together with the optimal inputs to control a temporal network.
Substituting the optimal inputs into the trajectory (\ref{solutionfortem}), we obtain the optimal trajectory.

\begin{figure}[H]
\centering
\includegraphics[width=0.85\textwidth]{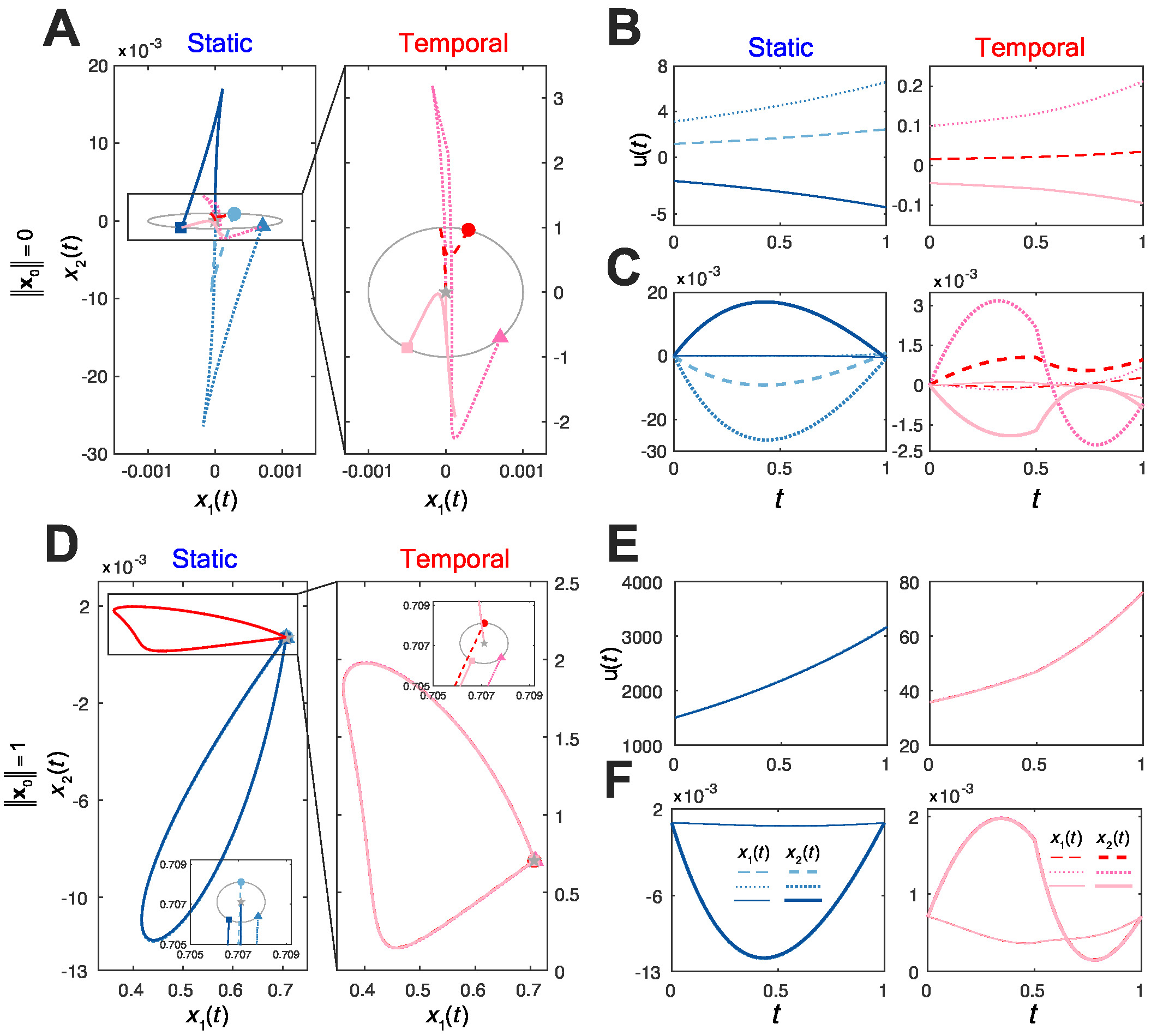}
\caption{\label{fig_SI_N2M2TrajUs}
\textbf{Control trajectory and inputs for static and temporal networks with a small control distance.}
(\textbf{A})-(\textbf{F}) correspond to a typical randomly-generated two-dimensional system (entries chosen randomly from $-1$ to $1$) with $M=2$ snapshots.
Three typical final states $\x _f$ are selected to illustrate the control trajectory with $\delta = \| \x _f - \x _0 \| = 10^{-3}$, where (A)-(C) represent the case $\| \x _0 \|= 0$ and (D) to (F) represent the case $\| \x _0 \| = 1$.
(A) gives the system trajectories for the network and its aggregated version from $\x _0 = 0$ (gray star) to three final states $\x _f$ (circle, triangle, and square) at an unit time, where its inputs (one driver node) and detailed trajectory of each node are shown in (B) and (C).
The insets therein enlarge the trajectories around the initial state.
(B) gives the inputs, which are in general non-smooth for temporal networks but smooth for temporal networks.
Note the disparate scales for the two cases, which implies the dramatic energy difference observed in the main text and Figs.~\ref{fig_SI_energyN10M2k4_minus3_minus1}-\ref{fig_SI_energy_N34Ndmore}.
(C) shows trajectories for each node, where the thin lines represent the first node $x_1(t)$, and thick lines the second node $x_2(t)$.
For (B) and (C), (E) and (F) the same color corresponds to the same final state $\x _f$ given in (B) and (D), respectively.
(D) gives the system trajectories for temporal and static networks from $\x _0 = (\sqrt{2}/2, \sqrt{2}/2)^\textrm{T}~\textrm{(gray star)}$ to $\x _f$ with $\x _f = \x _0 + \delta(0, 1)^\textrm{T} ~\textrm{(circle)},
~\x _0 +\delta(\sqrt{2}/2, -\sqrt{2}/2)^\textrm{T}~\textrm{(triangle)},
~\textrm{and}~\x _0 +\delta(-1/2, -\sqrt{3}/2)^\textrm{T}~\textrm{(square)}$.
In this case, the trajectories for different final states show no much difference since 
the control trajectories are qualitatively determined by the nature of the system dynamics, which in turn is determined by the orthant containing  $\x _0$.
In particular, the length of trajectories with different directions is almost the same (see Fig.~\ref{fig_SI_LdistN2M2M5}).
}
\end{figure}

\begin{figure}[H]
\centering
\includegraphics[width=1\textwidth]{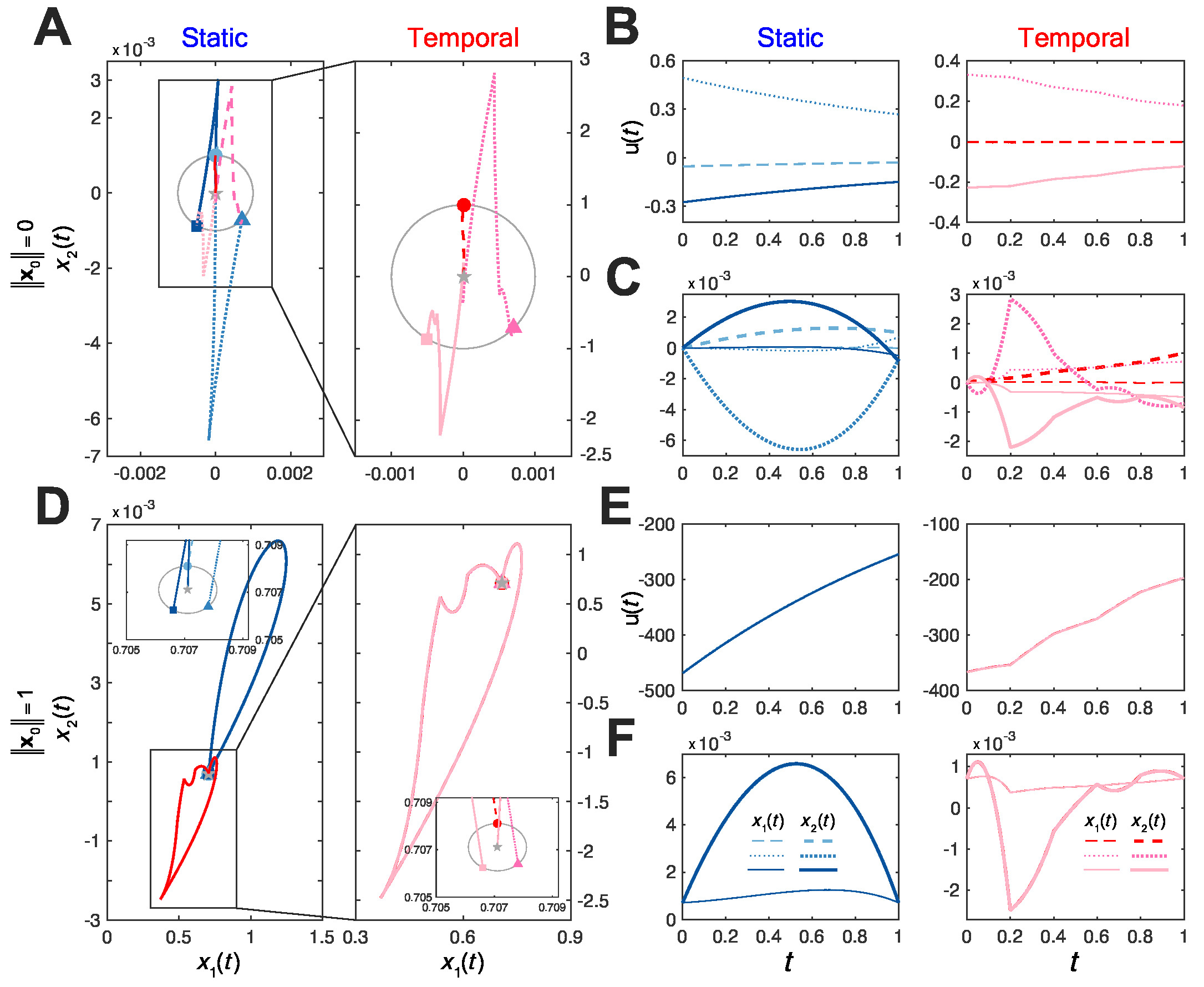}
\caption{\label{fig_SI_N2M5TrajUS}
\textbf{Control trajectories and inputs for static and temporal networks with a small control distance.}
Here we give the control trajectories and corresponding inputs for a random two-dimensional system with $M=5$ randomly-generated snapshots.
All other parameters are the same as those used in Fig.~\ref{fig_SI_N2M2TrajUs}.
}
\end{figure}

\begin{figure}[H]
\centering
\includegraphics[width=1\textwidth]{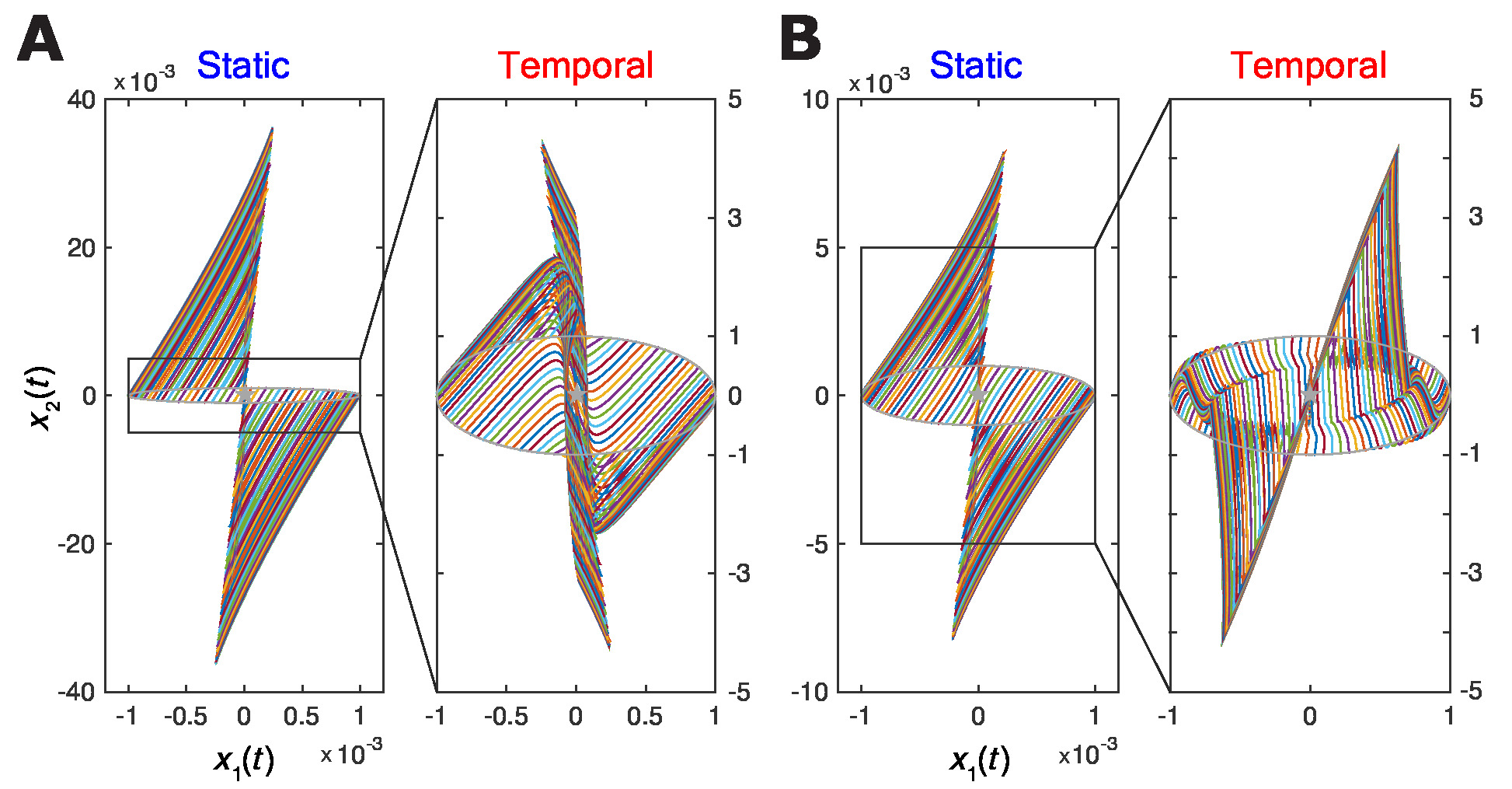}
\caption{\label{fig_SI_trajN2M2M5}
\textbf{Additional control trajectories for temporal and static networks with two and five snapshots.}
We select $100$ trajectories from $\x _0 = 0$ (indicated by a star) to $\x _f$ with $\| \x _f\| = 10^{-3}$ (\emph{i.e.} uniformly along the gray curve), for a randomly-generated temporal network and its static counterpart with (\textbf{A}) $N=2$, $M=2$, and (\textbf{B}) $N=2$, $M=5$.
All other parameters are the same as those used in Fig.~\ref{fig_SI_N2M2TrajUs}.
}
\end{figure}

\begin{figure}[H]
\centering
\includegraphics[width=1\textwidth]{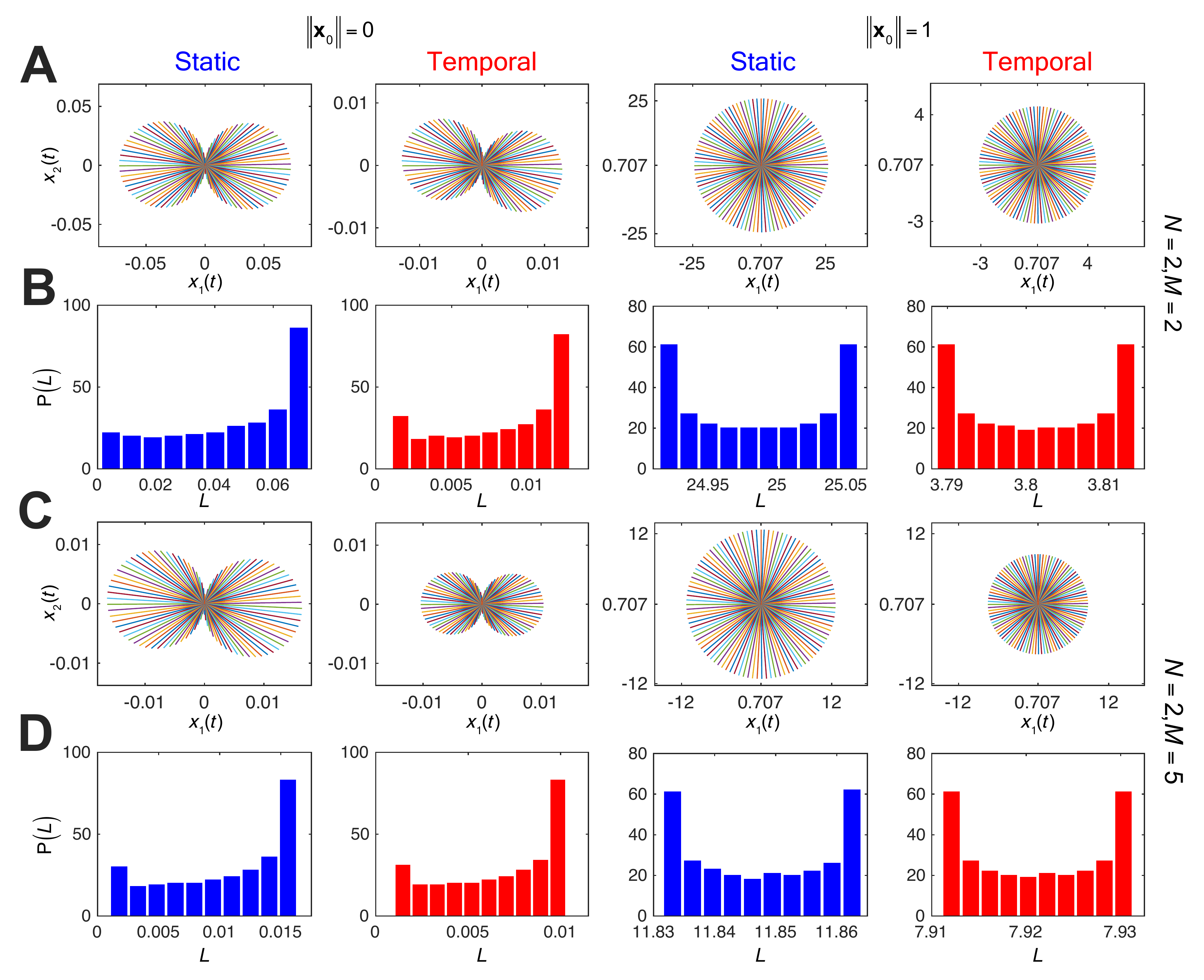}
\caption{\label{fig_SI_LdistN2M2M5}
\textbf{Distributions of control trajectory length for two dimensional systems.} 
The length of each trajectory is given in (\textbf{A}) and (\textbf{C}) for the cases of $\x _0 = \zero$ and $\x _0 \neq \zero$, and the corresponding distributions are shown in (\textbf{B}) and (\textbf{D}). 
We choose $100$ final states $\x _f$ uniformly at random from a circle of radius $\delta = 1$ centered on $\x _0$.
For a specific pair of initial and final states, the line is drawn from $\x _0$ toward $\x _f$ with the length $L$ of the corresponding control trajectory.
We find that the temporal networks decrease the length of a typical control trajectory regardless of whether or not $\x _0$ lies at the origin.
All other parameters are the same as those used in Fig.~\ref{fig_SI_N2M2TrajUs}.
}
\end{figure}

\begin{figure}[H]
\centering
\includegraphics[width=1\textwidth]{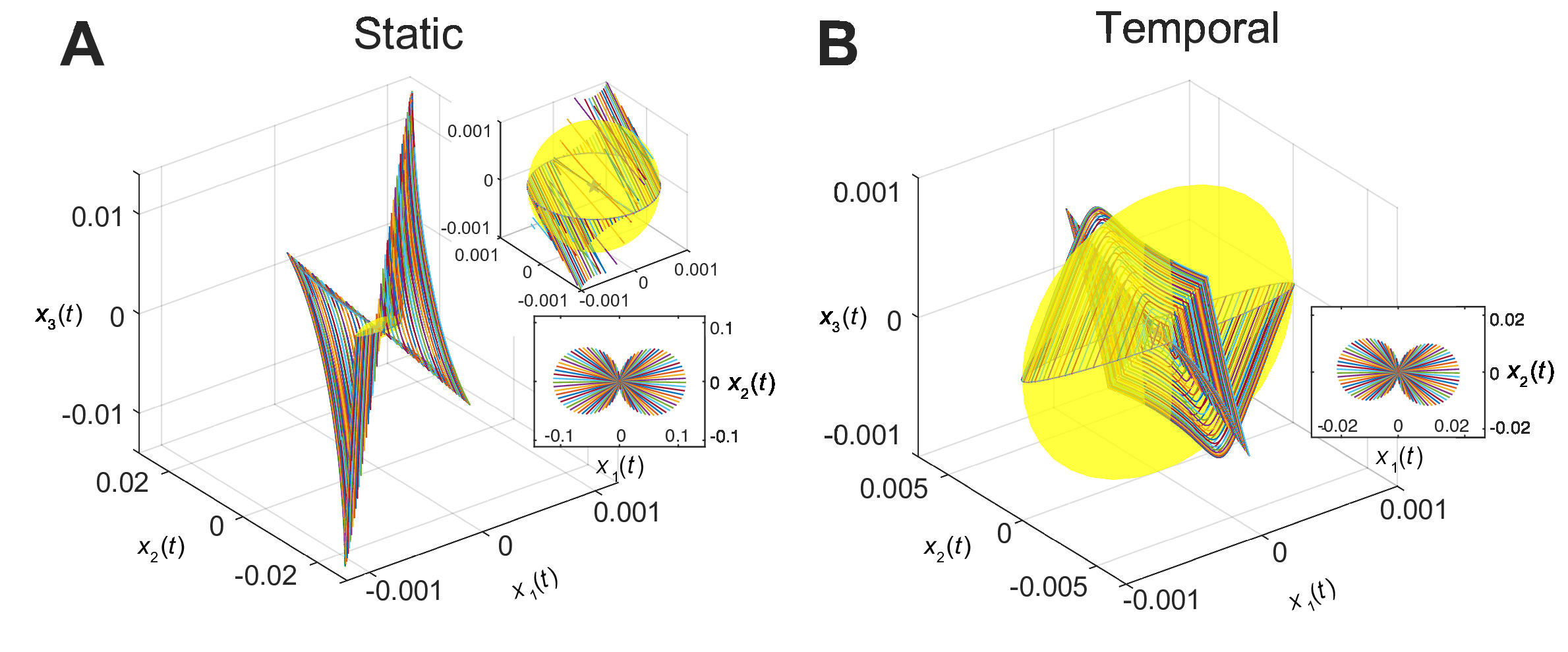}
\caption{\label{fig_SI_trajN3}
\textbf{Control trajectories in a three dimensional system.}
Each trajectory is from $\x _0$ (star) to a given $\x _f$ over $0 \leq t \leq 1$ for static (\textbf{A}) and temporal (\textbf{B}) networks. 
There are a total of $100$ randomly chosen $\x _f$ locating along a sphere (gray) centered on $\x _0$ with a radius of $\delta = 10^{-3}$.
We only plot those final states on equator for clarity.
Insets give the corresponding exact length of trajectories in each direction, which suggests that temporal networks reduce the length of trajectories significantly without changing its distribution (as that in Fig.~\ref{fig_SI_LdistN2M2M5}).
Here we consider three dimension systems and two snapshots for visualization, and the robustness of the results has been tested. 
All other parameters are the same as those used in Fig.~\ref{fig_SI_N2M2TrajUs}.
}
\end{figure}

In this section, we show control trajectories for temporal and static systems in two to three dimensions to give a visual understanding of the control non-locality of static networks.
For fixed control distance ($\delta = \|\x_f - \x_0 \| = 10^{-3}$),
there is a significant difference between control trajectories originating from $\x _0=\zero$ vs. other states 
(Fig.~\ref{fig_SI_N2M2TrajUs} and Fig.~\ref{fig_SI_N2M5TrajUS}).
When $\| \x _0 \| \neq 0$ ($\x _0 = (\sqrt{2}/2, \sqrt{2}/2)^\textrm{T}$), 
the length of the control trajectory is dominated by the need to make excursions to other orthants of the phase space, 
even for different final states 
at the same control distance 
(Fig.~\ref{fig_SI_N2M2TrajUs}B).
For Fig.~\ref{fig_SI_N2M2TrajUs}, Fig.~\ref{fig_SI_LdistN2M2M5}A, and Fig.~\ref{fig_SI_LdistN2M2M5}B we consider a temporal network with two nodes and two snapshots with different initial states $\x _0$ and control distance $\delta$.
With more snapshots, the results given above do not change, as shown in Fig.~\ref{fig_SI_N2M5TrajUS}, Fig.~\ref{fig_SI_LdistN2M2M5}C, and Fig.~\ref{fig_SI_LdistN2M2M5}D.
We also show the corresponding inputs and trajectory of each node.
Figure~\ref{fig_SI_trajN2M2M5} shows the full trajectories for an example two-dimensional system with two and five snapshots, for 100 different final states, and the distribution of the corresponding control trajectory lengths are shown in Fig.~\ref{fig_SI_LdistN2M2M5}.
Figure~\ref{fig_SI_trajN3} shows the same for a three dimensional system. 
In agreement with the results presented in Fig.~\ref{fig_main_4}, we find that the length of trajectories for temporal networks is always less than that for static networks, independent of the choice of $\x _0$ and the value of control distance.

We calculate $L$ numerically as
\begin{eqnarray*}
L &=& \int_{t_0}^{t_f} \| \dot{\x}(t) \| \mathrm{d}t
 = \int_{t_0}^{t_f} \sqrt{x_1'^2(t)+x_2'^2(t)+\cdots+x_N'^2(t)} \mathrm{d}t
\\ \nonumber & \approx &
\sum_{j=0}^{1/t_\text{step}} \sqrt{\sum_{i=1}^N \Big[x_i(t_j+t_\text{step}) - x_i(t_j)\Big]^2},
 \end{eqnarray*}
where $t_\text{step} = 0.025$ is the time step we choose to calculate $L$ numerically, 
and the limitation of the discrete approximation is exactly $L$ as $t_\text{step} \rightarrow 0$.

\begin{figure}[H]
\centering
\includegraphics[width=1\textwidth]{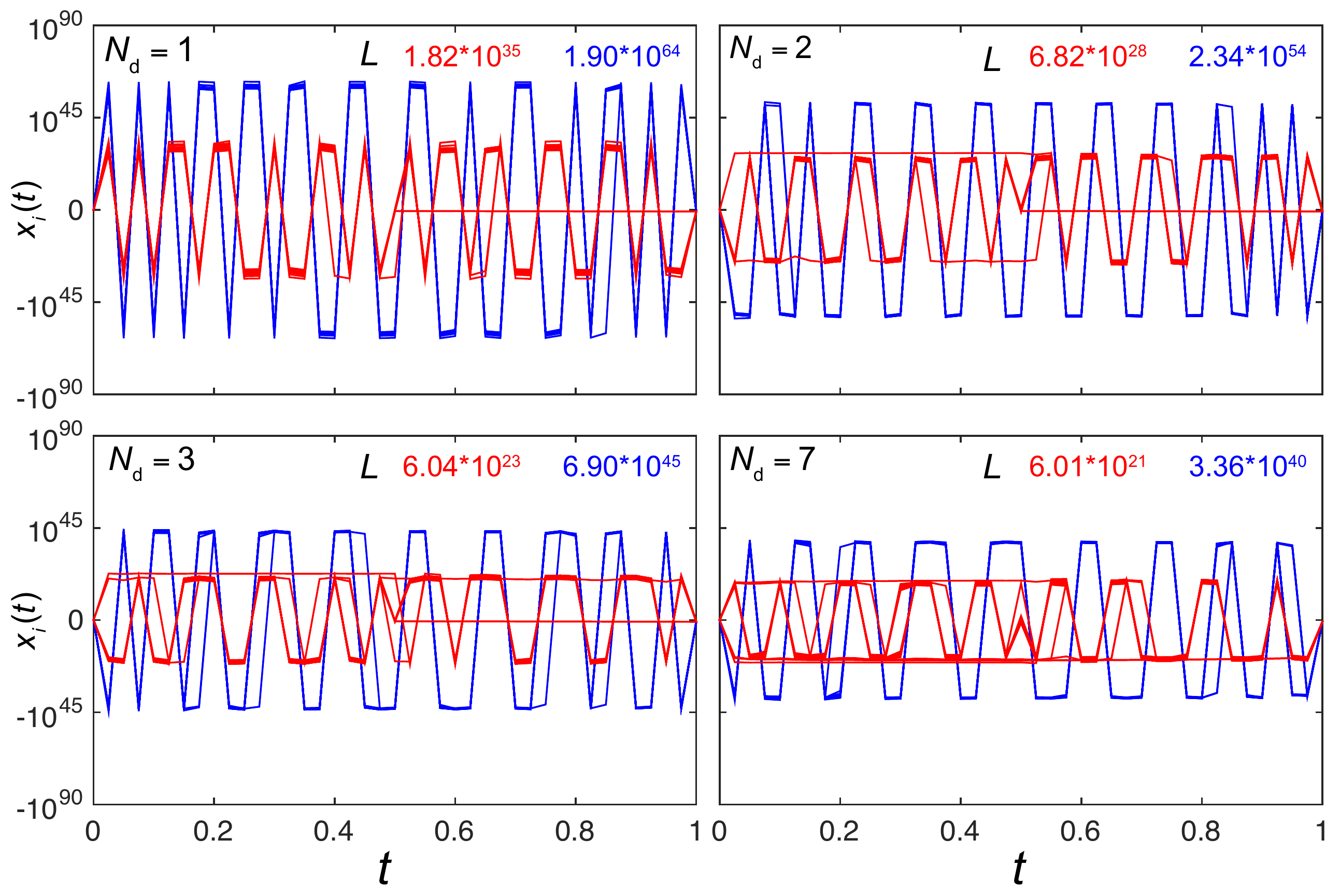}
\caption{\label{fig_SI_traj_N34Nd2}
\textbf{Locality of control trajectories in a real network.}
The panels show, for different numbers $N_d$ of driver nodes, the node states $x_i(t)$ as a function of time for control in the temporal (red) and static (blue) version of the ad hoc mobile communication network (1-ip6).
Here $\x _0=\zero$ and $\x _f$ is taken to be $(1,...,1)^\text{T}/ \sqrt{N}$ for $N = 34$ nodes.
The total length of the control trajectory $L$ is denoted in red (temporal) and blue (static).
We see that the true temporal version of this network exhibits considerably more local control trajectories than its aggregated counterpart, in line with the results shown for synthetic networks in Fig.~\ref{fig_main_4}C of the main text.
Moreover, $L$ decreases as $N_d$ increases for both the temporal and the static network.
Here $a_i = -1$ and $M=2$.
}
\end{figure}

In Fig.~\ref{fig_main_4}C for the technological network, we give the maximum magnitude of the state components (since the maximum state component dominates the whole length of the corresponding control trajectory), 
which can be expressed as
\begin{eqnarray*}
L_{i^*} &=& \max_{i} \sum_{j=0}^{1/t_\text{step}}  \Big | x_i(t_j+t_\text{step}) - x_i(t_j) \Big|.
 \end{eqnarray*}
Since $L_{i^*}$ is on the order of $10^{35}$ for the temporal network and $10^{64}$ for the static equivalent, $x_1(t)$ is sufficient to demonstrate that temporal networks exhibit more local trajectories for the technological network.
The state components of
the technological network in the cases of $1$, $2$, and $3$ driver nodes are shown in Fig.~\ref{fig_SI_traj_N34Nd2}.

\section{Supplementary Figures}

\end{document}